\documentclass[twocolumn,nofootinbib,superscriptaddress,floatfix,usenames,dvipsnames]{revtex4-2}\pdfoutput=1

\usepackage{amsmath}
\usepackage{amsfonts}
\usepackage[retainorgcmds]{IEEEtrantools}
\usepackage{tikz}
\usepackage{tikz-feynman}
\usepackage{float}
\usepackage[colorlinks=true, linktoc=page, citecolor=blue, urlcolor=blue]{hyperref}

\newcommand{\OO}{\mathcal{O}}
\newcommand{\x}{\textrm{x}}
\newcommand{\N}{\mathcal{N}}
\newcommand{\X}{\mathbb{X}}
\newcommand{\Y}{\mathbb{Y}}
\newcommand{\W}{\mathcal{W}}
\newcommand{\G}{\mathcal{G}}
\newcommand{\LL}{\mathcal{L}}
\newcommand{\C}{\mathcal{C}}
\newcommand{\CC}{\mathfrak{C}}
\newcommand{\I}{\mathcal{I}}
\newcommand{\J}{\mathcal{J}}
\newcommand{\F}{\mathfrak{F}}
\newcommand{\z}{\textrm{z}}

\hypersetup{pdfstartview={XYZ null null 0.75}}
\let\oldpageref\pageref
\renewcommand{\pageref}{\oldpageref*}

\begin{document}

\title{\textbf{Holographic correlators of semiclassical states in defect CFTs}}
\author{\textbf{George Georgiou}}
\email{ggeo@phys.uoa.gr}
\affiliation{Department of Physics, National and Kapodistrian University of Athens, 157 84, Athens, Greece}
\author{\textbf{Georgios Linardopoulos}}
\email{george.linardopoulos@wigner.hu}
\affiliation{Wigner Research Centre for Physics, Konkoly-Thege Mikl{ó}s \'{u}t 29-33, 1121 Budapest, Hungary}
\author{\textbf{Dimitrios Zoakos}}
\email{dzoakos@phys.uoa.gr}
\affiliation{Department of Physics, National and Kapodistrian University of Athens, 157 84, Athens, Greece}
\affiliation{Department of Engineering and Informatics, Hellenic American University, 436 Amherst St, Nashua, NH 03063 USA}
\begin{abstract}
\noindent We set up the computation of correlation functions for operators that are dual to semiclassical string states in strongly coupled defect conformal field theories (dCFTs). In the dCFT that is dual to the D3-D5 probe-brane system, we calculate the correlation function of two heavy operators perturbatively, in powers of the conformal ratio. We find that the leading term agrees with the prediction of the operator product expansion (OPE). In the case of two heavy BMN operators, we find agreement in subleading orders as well.
\end{abstract}

\maketitle
\section[Introduction]{Introduction \label{Section:Introduction}}
\noindent Deforming the gauge/string duality \cite{Maldacena97} by inserting probe branes on its string theory side \cite{KarchRandall01a, KarchRandall01b} has provided us with more and more realistic holographic models (AdS/dCFT correspondence) which are in principle solvable at strong coupling by string theory. Probe branes break many symmetries and supersymmetries of holographic theories, yet there is a single property that we would still like to keep. This property is planar integrability \cite{MinahanZarembo03, BenaPolchinskiRoiban03}. Integrability has the power of bridging the two opposing ends of holographic dualities (which are generally disconnected due to the weak/strong coupling dilemma), endowing holography with a genuine nonperturbative capacity \cite{Beisertetal12}. \\
\indent Integrability methods were introduced in the AdS/dCFT correspondence in 2015 \cite{deLeeuwKristjansenZarembo15}, sparking a wide range of weak-coupling computations at tree level \cite{Buhl-MortensenLeeuwKristjansenZarembo15, deLeeuwKristjansenMori16, deLeeuwKristjansenLinardopoulos18a, KristjansenMullerZarembo20a} and one-loop order \cite{Buhl-MortensenLeeuwIpsenKristjansenWilhelm16a, Buhl-MortensenLeeuwIpsenKristjansenWilhelm16c}. Asymptotic all-loop results appeared in \cite{Buhl-MortensenLeeuwIpsenKristjansenWilhelm17a, GomborBajnok20a, GomborBajnok20b, KristjansenMullerZarembo20b, KristjansenMullerZarembo21}. Classical string integrability was shown in \cite{DekelOz11b, LinardopoulosZarembo21}. While the majority of works so far concerns the D3-D5 probe-brane system, more integrable setups are currently known,\footnote{See the review articles \cite{deLeeuw19, deLeeuwIpsenKristjansenWilhelm17, Linardopoulos20} for more.} such as the D3-D7 \cite{deLeeuwKristjansenLinardopoulos16, GimenezGrauKristjansenVolkWilhelm19, deLeeuwGomborKristjansenLinardopoulosPozsgay19} and the D2-D4 probe-brane system \cite{KristjansenVuZarembo21, GomborKristjansen22, Linardopoulos22}. \\
\indent The D3-D5 system consists of a probe D5-brane embedded in the AdS$_5\times\text{S}^5$ background which is generated by $N$ D3-branes. The relative orientation of the D-branes in flat space is shown in table \ref{Table:D3D5system}. The D5-brane wraps an AdS$_4\times\text{S}^2$ geometry which is supported by $k$ units of abelian flux through S$^2$. The flux forces $k$ of the D3-branes to terminate on one side of the D5-brane.
\vspace{.2cm}\renewcommand{\arraystretch}{1.1}\setlength{\tabcolsep}{5pt}
\begin{table}[H]\begin{center}\begin{tabular}{|c||c|c|c|c|c|c|c|c|c|c|}
\hline
& $t$ & $x_1$ & $x_2$ & $x_3$ & $x_4$ & $x_5$ & $x_6$ & $x_7$ & $x_8$ & $x_9$ \\ \hline
\text{D3} & $\bullet$ & $\bullet$ & $\bullet$ & $\bullet$ &&&&&& \\ \hline
\text{D5} & $\bullet$ & $\bullet$ & $\bullet$ & & $\bullet$ & $\bullet$ & $\bullet$&&& \\ \hline
\end{tabular}\caption{D3-D5 brane orientation in flat space\label{Table:D3D5system}}\end{center}\end{table}
On the dual gauge theory side, we encounter a 4-dimensional dCFT. Two copies of $\N=4$ super Yang-Mills (SYM) theory with different gauge groups, $SU(N-k)$ and $SU(N)$ are separated by a codimension-1 defect \cite{DeWolfeFreedmanOoguri01}. Two-point functions in this theory have been studied in \cite{deLeeuwIpsenKristjansenVardinghusWilhelm17, Widen17, deLeeuwKristjansenLinardopoulosVolk23}. However, apart from the early supergravity calculations of one-point functions in \cite{NagasakiYamaguchi12,KristjansenSemenoffYoung12b}, the systematic computation of correlators in strongly coupled dCFTs with strings is still missing.\footnote{On the other hand, interesting results have been obtained with supersymmetric localization \cite{RobinsonUhlemann17, Wang20a, KomatsuWang20, BeccariaCaboBizet23}. Note also the holographic computations \cite{BakChenWu11, BissiKristjansenYoungZoubos11, CaputaMelloKochZoubos12, HiranoKristjansenYoung12, Lin12, KristjansenMoriYoung15} of correlation functions between open strings and finite-size branes such as giant gravitons.} The aim of the present letter is to fill this gap. First we address an important open problem in AdS/dCFT, that is the computation of two-point functions at strong coupling. Second, we provide a systematic framework for the calculation of correlation functions involving operators dual to semiclassical string states in strongly coupled AdS/dCFT. \\
\indent In section \ref{Section:ThreePointFunction} we compute the three-point function of two heavy BMN operators and a light BPS scalar at strong coupling in $SO(3)\times SO(3)$ symmetric $\N=4$ SYM. In section \ref{Section:OnePointFunction} we revisit the computation of the one-point function of a BPS operator in strongly coupled D3-D5 dCFT. In section \ref{Section:TwoPointFunction} we set up the computation of correlation functions for operators that are described by semiclassical worldsheets in the presence of a defect brane. To illustrate our method, we compute the two-point function of two heavy operators in strongly coupled D3-D5 dCFT. In section \ref{Section:OperatorProductExpansion} we compare our findings for the two-point function to the prediction of the OPE. We report complete agreement between the leading term of our strong coupling results and the leading term of the OPE for an arbitrary choice of heavy operators. For the case of two BMN operators, agreement is shown up to next-to-next-to-leading (NNLO) order.
\section[Three-point function]{Three-point function \label{Section:ThreePointFunction}}
\noindent Three and higher-point correlation functions can be computed in strongly coupled AdS/CFT when one of the operators is dual to a supergravity mode, based on a method that was developed in \cite{BerensteinCorradoFischlerMaldacena99} and applied to $\N=4$ SYM by \cite{Zarembo10c, CostaMonteiroSantosZoakos10}. Let $\W$ be a nonlocal operator of $\N = 4$ SYM (e.g.\ a Wilson loop or a product of local operators) that is dual to a classical string worldsheet and $\OO_I\left(y\right)$ a local operator of $\N = 4$ SYM that is dual to the scalar supergravity field $\phi_I\left(y,w\right)$. Defining
\begin{IEEEeqnarray}{l}
\left\langle\OO_I\left(y\right)\right\rangle_{\W} \equiv \frac{\left\langle\W \OO_I\left(y\right)\right\rangle_{\N = 4}}{\left\langle\W\right\rangle_{\N = 4}}, \label{CorrelationFunction1}
\end{IEEEeqnarray}
the correlator can be computed at strong coupling from
\begin{IEEEeqnarray}{ll}
\left\langle\OO_I\left(y\right)\right\rangle_{\W} = &\lim_{w\rightarrow 0}\bigg[\frac{\pi}{w^{\Delta_I}}\sqrt{\frac{2}{\Delta_I - 1}} \nonumber \\
&\big\langle\phi_I\left(y,w\right)\cdot \frac{1}{Z_{\text{str}}}\int D\X\,e^{-S_{\text{str}}\left[\X\right]}\big\rangle_{\text{bulk}}\bigg], \qquad \label{CorrelationFunction2}
\end{IEEEeqnarray}
where the prefactor ensures that the scalar field $\phi_I$ (scaling dimension $\Delta_I$) is properly normalized. Moreover, $S_{\text{str}}\left[\X\right]$ is the classical string action
\begin{IEEEeqnarray}{l}
S_{\text{str}} = -\frac{T_2}{2}\int d^2\sigma\sqrt{-\gamma}\gamma^{ab}\partial_a\X^M\partial_b\X^N g_{MN} + \ldots, \qquad \label{StringPolyakovAction}
\end{IEEEeqnarray}
in the absence of the fermions, the dilaton and the Kalb-Ramond field. In \eqref{StringPolyakovAction}, $T_2 \equiv \left(2\pi\alpha'\right)^{-1}$ is the string tension and $\X$ are the target-space coordinates, aka the embedding coordinates of the string worldsheet in AdS$_5\times\text{S}^5$. Parameter matching between the two sides of AdS$_5$/CFT$_4$ also leads to the identification $\lambda = \ell^4/\alpha'^{\,2}$, where $\lambda \equiv g_{\text{\scalebox{.8}{YM}}}^2 N$ is the 't Hooft coupling of $\N = 4$ SYM. \\
\indent The string action $S_{\text{str}}$ depends indirectly on the bulk supergravity modes $\phi_I$ via a disturbance that is induced on the fields of type IIB supergravity by a local operator insertion. In our case the relevant perturbations are:
\begin{IEEEeqnarray}{l}
g_{MN} = \hat{g}_{MN} + \delta g_{MN} \label{FieldPerturbationIIBa} \\
C_{MNPQ} = \hat{C}_{MNPQ} + \delta C_{MNPQ}, \label{FieldPerturbationIIBb}
\end{IEEEeqnarray}
where $g_{MN}$ is the graviton and $C_{MNPQ}$ is the 4-form Ramond-Ramond (RR) potential of type IIB supergravity. The corresponding background solution consists of the AdS$_5\times\text{S}^5$ metric $\hat{g}_{MN}$ (spelled out in \eqref{MetricAdS5xS5}) and the 4-form potential $\hat{C}_{MNPQ}$ (given in \eqref{RamondRamondPotentialAdS5xS5}). Both perturbations in \eqref{FieldPerturbationIIBa}--\eqref{FieldPerturbationIIBb} can be expressed as linear combinations of the bulk modes $\phi_I$ and their derivatives:
\begin{IEEEeqnarray}{l}
\delta g_{MN} = V^I_{MN} \cdot \phi_I, \quad \delta C_{MNPQ} = v^I_{MNPQ} \cdot \phi_I, \qquad \label{VertexOperatorsIIB}
\end{IEEEeqnarray}
where $V^I_{MN}$ and $v^I_{MNPQ}$ are differential operators which depend on the target-space coordinates $\X$. \\
\indent The string action is expanded around $\phi_I = 0$; in the strong coupling regime ($\lambda\to\infty$), the path integral in \eqref{CorrelationFunction2} is dominated by a saddle point which corresponds to classical solutions $\X_{\text{cl}}$ of the string equations of motion. In addition, $\langle\phi_I\left(y,w\right)\rangle = 0$ and $\left\langle\phi_I\left(y,w\right)\phi_I\left(x,z\right)\right\rangle$ is the bulk-to-bulk propagator \eqref{PropagatorBulkToBulk1} of the scalar field $\phi_I$ in Euclidean AdS$_5$. Taking the limit in \eqref{CorrelationFunction2}, we obtain the correlator to leading order in the perturbation $\phi_I$,
\begin{IEEEeqnarray}{ll}
\left\langle\OO_I\left(y\right)\right\rangle_{\W} = &-\frac{1}{4\ell^2}\sqrt{\frac{2\lambda}{\Delta_I - 1}} \int d^2\sigma\, \partial_a\X^M \partial^a\X^N \nonumber \\
& V^I_{MN}\left(\X,\partial_x,\partial_z\right) \G_{\Delta_I}\left(x,z;y\right) + \ldots, \qquad \label{CorrelationFunction3}
\end{IEEEeqnarray}
in the conformal gauge, $\gamma_{ab} = \text{diag}\left(-,+\right)$. The boundary limit $\G_{\Delta_I}$ of the bulk-to-bulk propagator \eqref{PropagatorBulkToBulk1} is \eqref{PropagatorBulkToBoundary}.

\vspace{.2cm}\begin{figure}[H]\begin{center}\begin{tikzpicture}
\draw (0,0) circle(2); \draw[line width=0.5mm, red] (-1.414,1.414) arc (250:290:4.15);
\node (O1) at (-2.3,1.6) {$\OO_1\left(\x_1,0\right)$};
\node (O2) at (2.3,1.6) {$\OO_2\left(\x_2,0\right)$};
\node (O3) at (-0.1,1.5) {$\left(x,z\right)$};
\node (O4) at (-0.1,-2.3) {$\OO_I\left(y,0\right)$};
\begin{feynman}
\vertex (a) at (-.1,1.15);
\vertex (b) at (0.1,-2);
\diagram*{(a) -- [photon, edge label=\(\phi_I\)] (b)};
\end{feynman}
\end{tikzpicture}\caption{Heavy-heavy-light (HHL) correlator}\label{Figure:ThreePointFunction}\end{center}\end{figure}
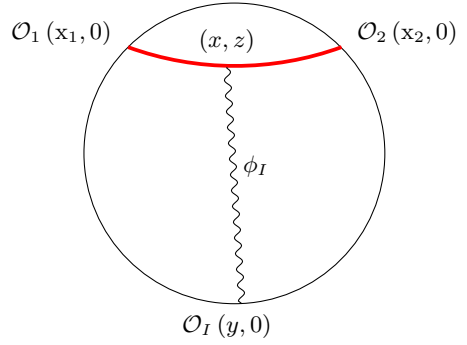
\indent Now take $\OO_I$ to be a chiral primary operator (CPO) of $\N = 4$ SYM with length $L$.\footnote{See appendix \ref{Appendix:ChiralPrimaryOperators} for the definition of CPOs in $\N = 4$ SYM.} Then the vertex operators that appear in \eqref{VertexOperatorsIIB} are given by \eqref{VertexOperators1}--\eqref{VertexOperators3}. Using \eqref{VertexOperators1}--\eqref{VertexOperators3} in \eqref{CorrelationFunction3}, we obtain an expression for the string correlation function \eqref{CorrelationFunction1}, \eqref{CorrelationFunction3} which simplifies significantly in the $y_i\rightarrow\infty$ limit \cite{Zarembo10c}:
\begin{IEEEeqnarray}{l}
\left\langle\OO_I^{\text{CPO}}\left(y\right)\right\rangle_{\W} = -\frac{L\sqrt{2(L-1)\lambda}}{4\pi^2 \N_L y^{2L}} \int d^2\sigma \, Y_I\left(x_{\mu}\right) \nonumber \\
\bigg\{-z^{L-2}\partial_a\X^i\partial^a\X^i + z^{L-2}\partial_a\X^z\partial^a\X^z + \nonumber \\
+ z^L\ell^{-2}\partial_a\X^{\mu}\partial^a\X^{\nu} \hat{g}_{\mu\nu}\bigg\}, \qquad i = 0,\ldots,3, \qquad \label{CorrelationFunction4}
\end{IEEEeqnarray}
where $\N_L$ is defined in \eqref{VertexOperators4}. Taking the operator $\W$ to be $\W \equiv \OO^{\dag}_{1}\OO_{2}$, where $\OO_{\mathfrak{i}}$ ($\mathfrak{i} = 1,2$) is a BMN chiral primary of length $L_{\mathfrak{i}}$,
\begin{IEEEeqnarray}{c}
\OO_{\mathfrak{i}} = \frac{1}{\sqrt{L_{\mathfrak{i}}}} \left(\frac{4\pi^2}{\lambda}\right)^{\frac{L_{\mathfrak{i}}}{2}} \text{tr}\left[Z^{L_{\mathfrak{i}}}\right], \quad Z \equiv \Phi_1 + i\Phi_2, \qquad \label{BMNoperators}
\end{IEEEeqnarray}
and $L = L_1 - L_2$ is small,\footnote{So that $\OO_1 \approx \OO_2$ and $\OO_I^{\text{CPO}}$ is a light operator.} the classical string solution that is holographically dual to $\W$ is given by \cite{Tsuji07, JanikSurowkaWereszczynski10}
\begin{IEEEeqnarray}{c}
x_3 = \bar\x + R \tanh\omega\tau, \qquad z = \frac{R}{\cosh\omega\tau} \qquad \label{BMNstring1a} \\
\psi = 0, \qquad \varphi = i\,\omega\tau, \qquad \theta = \frac{\pi}{2}, \qquad \label{BMNstring1b}
\end{IEEEeqnarray}
where $\omega = L_2/\sqrt{\lambda}$ and the 5-sphere parametrization can be found in \eqref{MetricS5so3so3}. The operators $\OO_{1,2}$ are located at the points $\x_{1,2}$ on the $x_3$ axis and a small distance from each other. In other words, $R = \x_{12}/2$ is small:
\begin{IEEEeqnarray}{l}
R = \frac{|\x_1 - \x_2|}{2} = \frac{\x_{12}}{2}, \qquad \bar\x = \frac{\x_1 + \x_2}{2}. \qquad \label{BMNstring2}
\end{IEEEeqnarray}
The three-point function is depicted in figure \ref{Figure:ThreePointFunction}. The red line represents the string worldsheet (heavy state) and the curly line represents the CPO (light state). \\
\indent Plugging the solution \eqref{BMNstring1a}--\eqref{BMNstring2} into \eqref{CorrelationFunction4}, we obtain the following expression for the three-point function $\left\langle\OO_I^{\text{CPO}}\left(y\right)\right\rangle_{\W}$ in the large-distance limit $y_i\rightarrow\infty$:
\begin{IEEEeqnarray}{ll}
\left\langle\OO_I^{\text{CPO}}\left(y\right)\right\rangle_{\W}^{\text{BMN}} = &\frac{(-1)^{L/2} \ell^2}{2^{L+\frac{3}{2}} N} \cdot L_2\sqrt{L(L+1)(L+2)} \nonumber\\
& B\Big(\frac{L}{2}+1,\frac{1}{2}\Big) \cdot \frac{\x_{12}^L}{y^{2L}}. \qquad \label{ThreePointFunctionBMN}
\end{IEEEeqnarray}
For reasons that will become apparent in \S\ref{Section:OperatorProductExpansion}, where we will verify the validity of the OPE for defect correlators, we have used the $SO(3)\times SO(3)$ invariant spherical harmonics (given by $\CC_{L/2}$ in \eqref{D5braneIntegral} below), for which $L=2j$. \\
\indent Inserting \eqref{ThreePointFunctionBMN} and the (generic CFT) 2-point function,
\begin{IEEEeqnarray}{ll}
\left\langle\W\left(\x_1,\x_2\right)\right\rangle_{\N = 4} = \langle\OO^{\dag}_{1}\left(\x_1\right)\OO_{2}\left(\x_2\right)\rangle_{\N = 4} = \frac{\delta_{12}}{\x_{12}^{L_1+L_2}} \qquad \label{TwoPointFunctionCFT}
\end{IEEEeqnarray}
into the definition \eqref{CorrelationFunction1} of the correlator $\left\langle\OO_I^{\text{CPO}}\left(y\right)\right\rangle_{\W}$, we may compare the result with the generic form of three-point functions in CFTs (for $\Delta_1 + \Delta_2 - \Delta_3 = L_1 + L_2 - L$, $\Delta_2 + \Delta_3 - \Delta_1 = 0$, $\Delta_3 + \Delta_1 - \Delta_2 = 2L$) and extract the HHL structure constant
\begin{IEEEeqnarray}{ll}
\C_{12}^{I}= \frac{(-1)^{L/2}L_2}{2^{L+\frac{3}{2}}N} \sqrt{L(L+1)(L+2)} \cdot B\Big(\frac{L}{2}+1,\frac{1}{2}\Big). \qquad \label{ThreePointFunctionStructureConstant}
\end{IEEEeqnarray}
The structure constant \eqref{ThreePointFunctionStructureConstant} is protected from receiving quantum corrections, i.e.\ it is the same from weak to strong coupling \cite{LeeMinwallaRangamaniSeiberg98}. Interesting works on holographic three-point functions (e.g.\ calculations involving twist operators and conserved currents) include \cite{RoibanTseytlin10, Georgiou10, Georgiou11, GeorgiouLeePark13, BajnokJanikWereszczynski14, BajnokJanik17a}.
\section[One-point function]{One-point function \label{Section:OnePointFunction}}
\noindent We will now briefly revisit the computation of one-point functions at strong coupling in dCFTs. We focus on the D3-D5 system. The action of the probe D5-brane is the sum of the Dirac-Born-Infeld (DBI) and the Wess-Zumino (WZ) term:
\begin{IEEEeqnarray}{ll}
S_{\text{D5}} = -\frac{T_5}{g_s}\int \Big[&d^6\zeta\sqrt{\det\left(G_{ab} + 2\pi\alpha'F_{ab}\right)} + \nonumber \\
&+ 2\pi\alpha' F\wedge C\Big], \qquad \label{D5braneAction}
\end{IEEEeqnarray}
where $T_5 \equiv \left(2\pi\right)^{-5}\alpha'^{-3}$ is the D5-brane tension, $g_s = g_{\text{\scalebox{.8}{YM}}}^2/4\pi$ is the string coupling constant, $G_{ab}$ is the pullback of the IIB graviton field $g_{MN}$ \eqref{FieldPerturbationIIBa} on the 5-brane,
\begin{IEEEeqnarray}{c}
G_{ab} \equiv \partial_a \Y^M \partial_b \Y^N g_{MN},
\end{IEEEeqnarray}
$F_{ab}$ is the field strength of the worldvolume gauge field and $C$ is the 4-form RR potential \eqref{FieldPerturbationIIBb} of type IIB supergravity. For the target space coordinates $\Y$ we set
\begin{IEEEeqnarray}{c}
h_{ab} \equiv \partial_a \Y^M \partial_b \Y^N \hat{g}_{MN} + 2\pi\alpha'F_{ab}, \quad h \equiv \det h_{ab}. \qquad \label{InducedMetric1}
\end{IEEEeqnarray}
The embedding of the probe brane inside AdS$_5\times\text{S}^5$ can be found by solving the equations of motion that arise from the action \eqref{D5braneAction}. As it turns out \cite{KarchRandall01b}, the D5-brane wraps an AdS$_4\times\text{S}^2$ geometry that is parametrized by
\begin{IEEEeqnarray}{c}
y_3 = \kappa \cdot w, \qquad \kappa \equiv \frac{\pi k}{\sqrt{\lambda}} \equiv \tan\alpha, \qquad \tilde{\psi} = 0, \qquad \label{D5braneEmbedding}
\end{IEEEeqnarray}
where $k$ are the units of magnetic flux through S$^2$:
\begin{IEEEeqnarray}{c}
\int_{\text{S}^2}\frac{F}{2\pi} = k, \qquad F = \frac{k}{2} \cdot d\cos\tilde{\theta}\wedge d\tilde{\varphi}, \label{WorldvolumeFlux}
\end{IEEEeqnarray}
$(y,w)$ are the AdS$_5$ coordinates in \eqref{MetricAdS5xS5} and the S$^5$ coordinates in \eqref{MetricS5so3so3} carry a tilde. The worldvolume coordinates of the D5-brane are $(\zeta_0,\ldots,\zeta_5) = (y_0,y_1,y_2,w,\tilde{\theta},\tilde{\varphi})$. \\
\indent One-point functions of dCFT operators that are dual to a supergravity mode $\phi_I$ can again be computed in strongly coupled AdS/dCFT by means of the recipe \eqref{CorrelationFunction2}:
\begin{IEEEeqnarray}{ll}
\left\langle\OO_I\left(x\right)\right\rangle_{\text{D5}} = \lim_{z\rightarrow 0}\bigg[&\frac{\pi}{z^{\Delta_I}}\sqrt{\frac{2}{\Delta_I - 1}} \cdot\big\langle\phi_I\left(x,z\right) \nonumber \\
&\frac{1}{Z_{\text{D5}}}\int D\Y\,e^{-S_{\text{D5}}\left[\Y\right]}\big\rangle_{\text{bulk}}\bigg], \qquad \label{OnePointFunctionsD5brane1}
\end{IEEEeqnarray}
where $\Delta_I$ is the scaling dimension of the scalar field $\phi_I$ and the D5-brane action $S_{\text{D5}}$ is given by \eqref{D5braneAction}. Once more, $S_{\text{D5}}$ depends indirectly on the bulk supergravity modes $\phi_I$ via a disturbance that is induced on the fields of type IIB supergravity by a local operator insertion. Expanding the D-brane action around $\phi_I = 0$ in the strong coupling regime ($\lambda\to\infty$), where the D5-brane path integral is dominated by a saddle point that corresponds to classical solutions of the DBI equations of motion $\Y_{\text{cl}}$, and making use of the fact that $\left\langle\phi_I\left(x,z\right)\phi_I\left(y,w\right)\right\rangle$ is the bulk-to-bulk propagator \eqref{PropagatorBulkToBulk1}, we arrive at \cite{GubserKlebanovPolyakov98, Witten98a, NagasakiYamaguchi12}:
\begin{IEEEeqnarray}{ll}
\left\langle\OO_I\left(x\right)\right\rangle_{\text{D5}} = &-\frac{\pi}{z^{\Delta_I}}\sqrt{\frac{2}{\Delta_I - 1}} \cdot \frac{T_5}{g_s}\int d^6\zeta (\delta \LL_{\text{DBI}} + \nonumber \\[6pt]
& \delta \LL_{\text{WZ}}) \cdot \G_{\Delta_I}\left(y,w;x\right), \qquad \label{OnePointFunctionsD5brane2}
\end{IEEEeqnarray}
by applying \eqref{FieldPerturbationIIBa}--\eqref{VertexOperatorsIIB}. Again, $\G_{\Delta_I}\left(y,w;x\right)$ is the boundary limit \eqref{PropagatorBulkToBoundary} of the bulk-to-bulk propagator \eqref{PropagatorBulkToBulk1} and

\vspace{.2cm}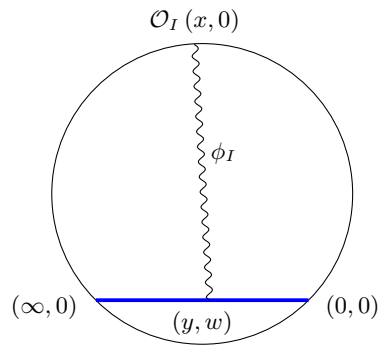
\begin{figure}[H]\begin{center}\begin{tikzpicture}
\draw (0,0) circle(2);\draw[line width=0.5mm, blue] (-1.414,-1.414) -- (1.414,-1.414);
\node (O3) at (-0.1,2.3) {$\OO_I\left(x,0\right)$};
\node (O4) at (2,-1.5) {$\left(0,0\right)$};
\node (O5) at (-2.1,-1.5) {$\left(\infty,0\right)$};
\node (O6) at (0,-1.72) {$\left(y,w\right)$};
\begin{feynman}
\vertex (a) at (-.1,2);
\vertex (b) at (0.1,-1.414);
\diagram*{(a) -- [photon, edge label=\(\phi_I\)] (b)};
\end{feynman}
\end{tikzpicture}\caption{One-point function of a CPO in defect CFT}\label{Figure:OnePointFunction}\end{center}\end{figure}
\begin{IEEEeqnarray}{l}
\delta \LL_{\text{DBI}} \equiv \sqrt{h} h^{ab} \partial_a \Y^M \partial_b \Y^N V^I_{MN}(\Y,\partial_y,\partial_w) \label{FluctuationsDBI} \\
\delta \LL_{\text{WZ}} \equiv 2\pi\alpha' \left(F\wedge v^I(\Y,\partial_y,\partial_w)\right). \label{FluctuationsWZ}
\end{IEEEeqnarray}
\indent Take $\OO_I$ to be a CPO of the D3-D5 dCFT with length $L$, situated at the point
\begin{IEEEeqnarray}{c}
x_0 = x_1 = x_2 = 0, \quad x_3 > 0,
\end{IEEEeqnarray}
i.e.\ at a distance $x_3$ from the planar boundary of $\N = 4$ SYM.\footnote{Setting $w = 0$ in \eqref{D5braneEmbedding}, it follows that the defect/boundary is just the plane $y_3 = 0$.} Of all the CPOs of $\N = 4$ SYM, only those which share the $SO(3)\times SO(3)$ global symmetry of the defect are expected to have nontrivial one-point functions. The same is true for their dual supergravity fields $s = s_I Y_I$ which can only depend on those spherical harmonics on S$^5$ which are $SO(3)\times SO(3)$ invariant (see e.g.\ \cite{NagasakiYamaguchi12} for a brief description). The $SO(3)\times SO(3)$ spherical harmonics depend on a single quantum number $j$ which is related to the length of the associated CPO via $L = 2j$. The one-point function is depicted in figure \ref{Figure:OnePointFunction}. The blue line represents the worldvolume of the D5-brane and the curly line is the CPO. \\
\indent Inserting the vertex operators \eqref{VertexOperators1}--\eqref{VertexOperators3} and the D5-brane parametrization \eqref{D5braneEmbedding}--\eqref{WorldvolumeFlux} into the one-point function formula \eqref{OnePointFunctionsD5brane2}--\eqref{FluctuationsWZ}, we are led to
\begin{IEEEeqnarray}{c}
\left\langle\OO_I^{\text{CPO}}\left(x_3\right)\right\rangle_{\text{D5}} = \frac{\C_{I}}{x_3^L}. \label{OnePointFunctionsDefect}
\end{IEEEeqnarray}
The one-point function structure constant $\C_{I}$ reads \cite{NagasakiYamaguchi12}:
\begin{IEEEeqnarray}{c}
\C_{I} = \frac{(-1)^{L/2}\sqrt{\lambda}}{\pi^{3/2}}\sqrt{\frac{L+2}{2L(L+1)}}\cdot \frac{\Gamma\left(L + \frac{1}{2}\right)}{\Gamma\left(L\right)} \I_{L-2,L+\frac{1}{2}}, \qquad \label{OnePointFunctionStructureConstantD5}
\end{IEEEeqnarray}
for even $L = 2j$ and nonnegative integer $j = 0,1,\ldots$ The analytic computation of the integral
\begin{IEEEeqnarray}{c}
\I_{a,b}\left(\kappa\right) \equiv \int\displaylimits_{0}^{\infty} du \, \frac{u^a}{\left[u^2 + (1 - \kappa u)^2\right]^b}, \label{IntegralsI}
\end{IEEEeqnarray}
for $a = L-2$, $b = L + 1/2$ can be found in appendix \ref{Appendix:Integrals}.
\section[Two-point function]{Two-point function \label{Section:TwoPointFunction}}
\noindent Two and higher-point correlation functions can still be computed to leading order in strongly coupled AdS/dCFT by the recipe \eqref{CorrelationFunction2}, \eqref{OnePointFunctionsD5brane1}. Let $\W$ be a nonlocal operator of $\N = 4$ SYM (Wilson loop, product of local operators, etc.) that is dual to a classical string worldsheet. Suppose that there is a probe D5-brane in the bulk of AdS$_5\times\text{S}^5$ which interacts with the semiclassical string state via a scalar type IIB supergravity mode $\phi_I$ whose scaling dimension is $\Delta_I$ and its mass is $m$. The supergravity mode is essentially emitted from the D-brane and absorbed by the semiclassical worldsheet. Then the ratio of the correlation function $\left\langle\W\right\rangle_{\text{D5}}$ in D5-brane deformed $\N = 4$ SYM over its value $\langle\W\rangle_{\N = 4} $ in pure $\N = 4$ SYM will be given at strong coupling by \cite{BerensteinCorradoFischlerMaldacena99}:
\begin{IEEEeqnarray}{ll}
\langle\widetilde{\W}\rangle_{\text{D5}} &\equiv \frac{\langle \W\rangle_{\text{D5}}}{\langle \W\rangle_{\N = 4}}= \big\langle\frac{1}{Z_{\text{str}}}\int D\X\,e^{-S_{\text{str}}\left[\X\right]} \nonumber \\
& \frac{1}{Z_{\text{D5}}}\int D\Y\,e^{-S_{\text{D5}}\left[\Y\right]}\big\rangle_{\text{bulk}}. \qquad \label{DefectCorrelator1}
\end{IEEEeqnarray}
This formula should in principle include all possible virtual states that can be exchanged between the string and the brane, i.e.\ CPOs and non-protected heavy string states. Only the former are taken into account in \eqref{DefectCorrelator1}. To determine the contribution of the latter, we should find the minimal surface $\mathcal{A}$ (with prescribed boundary conditions on the boundary of the AdS) that terminates on the brane. However, such states are exponentially suppressed as $e^{-\sqrt{\lambda}\mathcal{A}}$ compared to the CPOs \cite{KazamaKomatsu11}. Moreover, their contribution to the two-point correlators that we are considering below will be extremely suppressed (in the limit where the operator insertions are far from the brane) due to the large anomalous dimensions these operators acquire at strong coupling. \\

\vspace{.2cm}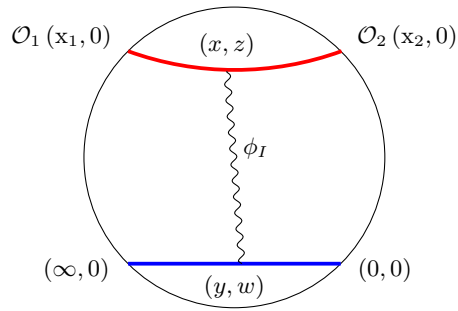
\begin{figure}[H]\begin{center}\begin{tikzpicture}
\draw (0,0) circle(2); \draw[line width=0.5mm, red] (-1.414,1.414) arc (250:290:4.15);\draw[line width=0.5mm, blue] (-1.414,-1.414) -- (1.414,-1.414);
\node (O1) at (-2.3,1.6) {$\OO_1\left(\x_1,0\right)$};
\node (O2) at (2.3,1.6) {$\OO_2\left(\x_2,0\right)$};
\node (O3) at (-0.1,1.5) {$\left(x,z\right)$};
\node (O4) at (2,-1.5) {$\left(0,0\right)$};
\node (O5) at (-2.1,-1.5) {$\left(\infty,0\right)$};
\node (O6) at (0,-1.72) {$\left(y,w\right)$};
\begin{feynman}
\vertex (a) at (-.1,1.15);
\vertex (b) at (0.1,-1.414);
\diagram*{(a) -- [photon, edge label=\(\phi_I\)] (b)};
\end{feynman}
\end{tikzpicture}\caption{Two-point function of BMN operators in D3-D5}\label{Figure:TwoPointFunction}\end{center}\end{figure}
\indent Expanding the string and the brane action around $\phi_I = 0$ as before and making use of the fact that in the strong coupling regime ($\lambda\to\infty$) both path integrals in \eqref{DefectCorrelator1} will be dominated by their saddle points (corresponding to the classical solutions $\X_{\text{cl}}$, $\Y_{\text{cl}}$), we obtain the defect correlation function of the operator $\W$ that is dual to an AdS$_5\times\text{S}^5$ semiclassical string state:
\begin{IEEEeqnarray}{ll}
\langle\widetilde{\W}\rangle_{\text{D5}} = &1 + \frac{T_2T_5}{2g_s} \int d^2\sigma \, d^6\zeta \bigg\{\delta\LL_{\text{str}}\left(\sigma,x,z\right) \nonumber \\
& \delta\LL_{\text{D5}}\left(\zeta,y,w\right) G_{\Delta_I}\left(x,z;y,w\right)\bigg\}, \qquad \label{DefectCorrelator2}
\end{IEEEeqnarray}
where $G_{\Delta_I}$ is the bulk-to-bulk propagator \eqref{PropagatorBulkToBulk1} of a scalar field (mass $m$, scaling dimension $\Delta_I$) in AdS$_5$ and
\begin{IEEEeqnarray}{ll}
\delta\LL_{\text{str}} = &\partial_a\X^M\partial^a\X^N\,V^I_{MN}\left(\X,\partial_x,\partial_z\right) + \ldots \qquad \label{FluctuationsString} \\
\delta\LL_{\text{D5}} = &\sqrt{h} h^{ab} \partial_a \Y^M \partial_b \Y^N V^I_{MN}(\Y,\partial_y,\partial_w) + \nonumber \\
& + 2\pi\alpha' \left(F\wedge v^I(\Y,\partial_y,\partial_w)\right). \qquad \label{FluctuationsD5brane}
\end{IEEEeqnarray}
\indent Let the supergravity mode $\phi_I$ be dual to a CPO $\OO^{\text{CPO}}_I$ of the D3-D5 dCFT with length $L$. The CPO and its dual supergravity field $s = s_I Y_I$ should share the $SO(3)\times SO(3)$ global symmetry of the defect. The S$^5$ spherical harmonics should then be $SO(3)\times SO(3)$ invariant and depend on a single quantum number $j$ which again fixes the length of the CPO to $L = 2j$. For simplicity let us also assume that the string worldsheet lies very close to the AdS boundary, that is $z\rightarrow 0$. The near-boundary expansion of the bulk-to-bulk propagator is given by:
\begin{IEEEeqnarray}{ll}
G_L\left(x,z;y,w\right) = \frac{L - 1}{2\pi^2}\cdot\bigg\{&1 + \frac{L\Lambda_w z^2}{\left(L - 1\right) K_w^2} + \nonumber \\
& + \OO\left(z^4\right)\bigg\}\cdot \left(\frac{z w}{K_w}\right)^L, \qquad \label{PropagatorBulkToBulk2}
\end{IEEEeqnarray}
where $K_w \equiv w^2 + \left(x - y\right)^2$ and $\Lambda_w \equiv 2 w^2 - \left(L - 1\right)\left(x - y\right)^2$ (since $\nu = L-2$, see appendix \ref{Appendix:ChiralPrimaryOperators}).
To integrate over the D5-brane coordinates $\zeta$ in \eqref{DefectCorrelator2}, we follow the same steps that we followed above for the computation of the defect one-point function \eqref{OnePointFunctionsD5brane2}--\eqref{OnePointFunctionStructureConstantD5}. We first compute the integrand by applying the vertex operators \eqref{VertexOperators1}--\eqref{VertexOperators3} on the propagator \eqref{PropagatorBulkToBulk2}. We are led to
\begin{IEEEeqnarray}{l}
\int d^6\zeta \, \delta\LL_{\text{D5}} \, G_L = -\frac{16\pi^{1/2}\CC_{L/2}\ell^6 L\left(L-1\right)}{\N_L} \nonumber \\
\sum_{n=0}^{\infty} \F_n \cdot \frac{z^{L+2n}}{x_3^{L+2n}}, \qquad \CC_{L/2} = \left(-\frac{1}{2}\right)^{\frac{L}{2}}\sqrt{\frac{L+2}{2L+2}}, \qquad \label{D5braneIntegral}
\end{IEEEeqnarray}
where the first two coefficients read, for $L = 2j$:
\begin{IEEEeqnarray}{ll}
\F_0 &= \frac{\Gamma\left(2j + \frac{1}{2}\right)}{\Gamma\left(2j + 2\right)} \cdot \I_{2j-2,2j+\frac{1}{2}} = \nonumber \\
&= \frac{\sqrt{\pi}\,\kappa^{2j+1}}{2j(2j+1)}\cdot\bigg\{1 + \frac{j(2j+1)}{2(2j-1)}\cdot\frac{1}{\kappa^2} + \ldots\bigg\} \\[6pt]
\F_1 &= \frac{-1}{2(2j-1)}\Bigg[\frac{\Gamma\left(2j+\frac{1}{2}\right)}{\Gamma(2j+2)} \cdot \I_{2j-2,2j+\frac{1}{2}} + \nonumber \\
&+ 2(2j+2)\cdot \frac{\Gamma\left(2j+\frac{3}{2}\right)}{\Gamma(2j+3)}\Big[2\kappa \cdot \I_{2j-1,2j+\frac{3}{2}} + \nonumber \\
&+ (2j-1)\cdot \I_{2j-2,2j+\frac{3}{2}}\Big] - 2(2j+2)(2j+3) \cdot \nonumber \\
& \frac{\Gamma\left(2j+\frac{5}{2}\right)}{\Gamma(2j+4)} \cdot \I_{2j,2j+\frac{5}{2}}\Bigg] = \nonumber \\
&= -\frac{\sqrt{\pi}\,\kappa^{2j+1}}{4(2j-1)} \cdot \bigg\{1 + \frac{j(2j+1)}{2(2j-1)}\cdot\frac{1}{\kappa^2} + \ldots \bigg\}. \qquad
\end{IEEEeqnarray}
We have also computed $\F_2$ but it's far too lengthy to be included here. The integrals $\I_{a,b}$ are known as power series of $\kappa \rightarrow \infty$ (see appendix \ref{Appendix:Integrals}). Note however that the ratios of the coefficients $\F_n$ depend only on $j$:
\begin{IEEEeqnarray}{ll}
{\F_1 \over \F_0} = -{j(2j+1) \over 2(2j-1)}, \quad {\F_2 \over \F_0} = {(j+1)(2j+1)(2j+3) \over 16(2j-1)}. \qquad \label{RatiosF}
\end{IEEEeqnarray}
\indent To obtain the value of the correlation function \eqref{DefectCorrelator2} we must also integrate over the string worldsheet coordinates. The integrand is again obtained by applying the vertex operators \eqref{VertexOperators1} on the D5-brane integral \eqref{D5braneIntegral}:
\begin{IEEEeqnarray}{l}
\delta\LL_{\text{str}}\left(\sigma,x,z\right) \int d^6\zeta \, \delta\LL_{\text{D5}}\left(\zeta,y,w\right) G_L\left(x,z;y,w\right). \qquad
\end{IEEEeqnarray}
Putting together the two contributions, we obtain the general form of the defect correlator \eqref{DefectCorrelator1}:
\begin{widetext}\begin{IEEEeqnarray}{ll}
\langle\widetilde{\W}\rangle_{\text{D5}} = 1 &+ \frac{(-1)^L (L+2) \lambda}{16N\pi^{5/2}} \int_{0}^{2\pi}\int_{-\infty}^{+\infty} d\sigma \, d\tau \cdot \sum_{n=0}^{\infty} \F_n \cdot \frac{z^{L+2n}}{x_3^{L+2n}} \bigg\{\Big[\left(L^2 + L + 4n\right)\left(\partial_a\X^i\partial^a\X^i\right) - \nonumber \\
& - \left(L^2 + (8n+1)L + 8n^2\right)\left(\partial_a\X^z\partial^a\X^z\right)\Big] z^{-2} - L\left(L+1\right)\left(\ell^{-2}\partial_a\X^{\mu}\partial^a\X^{\nu} \hat{g}_{\mu\nu}\right) + \nonumber \\
& + 4\left(L + 2n\right)\left(L + 2n +1\right)\left(\partial_a\X^3\partial^a\X^z\right)x_3^{-1}z^{-1} - 2 \left(L + 2n\right)\left(L + 2n +1\right)\left(\partial_a\X^3\partial^a\X^3\right)x_3^{-2}\bigg\}. \qquad \label{DefectCorrelator3}
\end{IEEEeqnarray}\end{widetext}

\indent For arbitrary heavy semiclassical operators, the leading term ($n=0$) in the correlator \eqref{DefectCorrelator3} factorizes into the product of the correlator \eqref{CorrelationFunction4} and the one-point function \eqref{OnePointFunctionsDefect} as follows (for $y_i \rightarrow \infty$ and $x_3 = \x_2 = \text{const.}$):
\begin{IEEEeqnarray}{l}
\langle\widetilde{\W}\rangle_{\text{D5}} = 1 + \left\langle\OO_I^{\text{CPO}}\left(\x_2\right)\right\rangle_{\text{D5}} \left\langle\OO_I^{\text{CPO}}\left(y\right)\right\rangle_{\W} y^{2L} + \ldots \qquad \label{ArbitraryCorrelator}
\end{IEEEeqnarray}
Non-protected operators that are exchanged between the heavy states and the D5-brane acquire very large dimensions at strong coupling and contribute only to subleading orders. For two heavy operators $\OO_{1,2}$, \eqref{ArbitraryCorrelator} becomes:
\begin{IEEEeqnarray}{l}
\frac{\langle\OO_{1}\left(\x_1\right)\OO_{2}\left(\x_2\right)\rangle_{\text{D5}}}{\langle\OO_{1}\left(\x_1\right)\OO_{2} \left(\x_2\right)\rangle_{\N = 4}} = 1 + 2^L \C_I \, \textrm{C}_{12}^I \, \xi^j + \OO({\xi^{j+1}}), \qquad \label{TwoPointFunctionArbitrary1}
\end{IEEEeqnarray}
where $\C_I$ is the one-point function structure constant \eqref{OnePointFunctionStructureConstantD5} and $\textrm{C}_{12}^I$ is the HHL structure constant. Moreover $L = L_1 - L_2 = 2j$ ($L_{1,2}$ are the lengths of the operators) and
\begin{IEEEeqnarray}{ll}
\xi \equiv \frac{\x_{12}^2}{4\x_1\x_2} \equiv \frac{v^2}{1-v^2}, \qquad \x_{12} \equiv |\x_1 - \x_2|, \qquad
\end{IEEEeqnarray}
defines the conformal ratios. We will see in \S\ref{Section:OperatorProductExpansion} below that the leading order behavior \eqref{TwoPointFunctionArbitrary1} is in complete agreement with the OPE. For two BMN chiral primary operators \eqref{BMNoperators}, agreement will also be shown for the subleading terms. Let us first compute their two-point function. \\
\indent Take $\W$ to be the operator $\W \equiv \OO^{\dag}_{1}\OO_{2}$, where $\OO_{\mathfrak{i}}$ ($\mathfrak{i} = 1,2$) are BMN chiral primaries \eqref{BMNoperators} that are located at the points $\x_{1,2}$ on the $x_3$ axis and a small distance $\x_{12}$ from each other (see figure \ref{Figure:TwoPointFunction}). As we have already mentioned in \S\ref{Section:ThreePointFunction}, when $L = L_1 - L_2$ is small, $\W$ is holographically dual to the classical (pointlike) string solution \eqref{BMNstring1a}--\eqref{BMNstring2} with $R = \x_{12}/2 \rightarrow 0$. In addition, the two heavy operators $\OO_{1,2}$ are nearly equal ($\OO_1 \approx \OO_2$) and $\OO_I^{\text{CPO}}$ is a light operator. Using the identification \eqref{BMNstring2} we may write the conformal ratios $\xi$ and $v$ as
\begin{IEEEeqnarray}{ll}
\xi \equiv \frac{\x_{12}^2}{4\x_1\x_2} = \frac{R^2}{\bar\x^2 - R^2} \Rightarrow \frac{R^2}{\bar\x^2} = \frac{\xi}{\xi + 1} \equiv v^2. \qquad \label{ConformalRatios}
\end{IEEEeqnarray}
Inserting the ansatz \eqref{BMNstring1a}--\eqref{BMNstring2} into the formula \eqref{DefectCorrelator3} for the defect two-point function $\langle\widetilde{\W}\rangle_{\text{D5}}$ we arrive at
\begin{widetext}\begin{IEEEeqnarray}{ll}
\frac{\langle\OO^{\dag}_{1}\left(\x_1\right)\OO_{2}\left(\x_2\right)\rangle_{\text{D5}}}{\langle\OO^{\dag}_{1}\left(\x_1\right)\OO_{2} \left(\x_2\right)\rangle_{\N = 4}} = 1 &- \frac{(-1)^L(L+2)\omega\lambda}{8N\pi^{3/2}}\sum_{n=0}^{\infty} \F_n \cdot \bigg\{2\Big[L^2 + 2n + L - v^{-2}(L + 2n)(L + 2n + 1)\Big]\J_{\frac{L}{2} + n, L + 2n + 2} + \nonumber \\
& + 8n(L+n)\Big[2v^{-1}\J_{\frac{L}{2} + n - 1, L + 2n + 1} - \left(v^{-2} - 1\right)\J_{\frac{L}{2} + n - 1, L + 2n + 2}\Big]\bigg\}, \qquad \label{TwoPointFunctionBMN1}
\end{IEEEeqnarray}\end{widetext}
where the integrals $\J_{a,b}$ are known as power series of $v \rightarrow 0$ (see appendix \ref{Appendix:Integrals}). Plugging \eqref{IntegralJ1}--\eqref{IntegralJ3} into the two-point function \eqref{TwoPointFunctionBMN1} we find, for $L=2j$:
\begin{IEEEeqnarray}{l}
\frac{\langle\OO^{\dag}_{1}\left(\x_1\right)\OO_{2}\left(\x_2\right)\rangle_{\text{D5}}}{\langle\OO^{\dag}_{1}\left(\x_1\right)\OO_{2} \left(\x_2\right)\rangle_{\N = 4}} = 1 + \frac{2j^2(j+1)L_2\sqrt{\lambda}}{N\pi^{3/2}} \, B\left(j,1/2\right) \nonumber \\
\F_0 \, \xi^{j}\Bigg\{1 + \frac{2j}{(2j+1)}\frac{\F_1}{\F_0}\xi + \frac{4j(j+1)}{(2j+1)(2j+3)}\frac{\F_2}{\F_0} \xi^2 + \ldots \Bigg\}. \nonumber \\ \label{TwoPointFunctionBMN2}
\end{IEEEeqnarray}
Taking into account the ratios \eqref{RatiosF}, our finding (checked up to NNLO) is in perfect agreement with our expectations from the operator product expansion (OPE) as we show right below. It is quite straightforward to obtain the two-point function to any subsequent perturbative order. Complete agreement with the OPE is expected.
\section[Operator product expansion]{Operator product expansion \label{Section:OperatorProductExpansion}}
\noindent In the present section we show that the leading-order defect two-point function \eqref{TwoPointFunctionArbitrary1} of two arbitrary heavy operators and the NNLO defect two-point function of two BMN chiral primaries \eqref{TwoPointFunctionBMN2} agree with the OPE. The bulk channel OPE reads:
\begin{IEEEeqnarray}{l}
\OO_1\left(\x_1\right)\OO_2\left(\x_2\right) = \frac{\delta_{12}}{\x_{12}^{\Delta_1 + \Delta_2}} + \nonumber \\
+ \sum_{I}\frac{\textrm{C}_{12}^{I}}{\x_{12}^{\Delta_1 + \Delta_2 - \Delta_I}} \cdot C\left[\x_1 - \x_2,\partial_{\x_2}\right]\OO_I\left(\x_2\right), \qquad \label{OperatorProductExpansion1}
\end{IEEEeqnarray}
where $C$ is a differential operator, $\Delta_{1,2}$, $\Delta_I$ are the dimensions of the operators $\OO_{1,2}$, $\OO_I$, and $\textrm{C}_{12}^{I}$ their CFT three-point function. The OPE \eqref{OperatorProductExpansion1} is valid independently of the presence of defects. Inserting \eqref{OperatorProductExpansion1} into the general formula for the defect two-point function
\begin{IEEEeqnarray}{l}
\left\langle\OO_1\left(\z_1,\textbf{x}_1\right)\OO_2\left(\z_2,\textbf{x}_2\right)\right\rangle = \frac{f_{12}\left(\xi\right)}{\left|\z_1\right|^{\Delta_1}\left|\z_2\right|^{\Delta_2}} \qquad \label{TwoPointFunctionsDefect}
\end{IEEEeqnarray}
and using the generic form of one-point functions \eqref{OnePointFunctionsDefect} (for $\Delta_I = L$) we obtain,\footnote{The defect is located at $\z = 0$ and $\x_{\mathfrak{i}} = \left(\z_{\mathfrak{i}}, \textbf{x}_{\mathfrak{i}}\right)$, for $\mathfrak{i} = 1,2$. The extra factor $2^{\Delta_I}$ in \eqref{StructureConstantDefect} compensates for the missing $2x_3$ in the denominator of \eqref{OnePointFunctionsDefect}, cf.\ \cite{McAvityOsborn95, LiendoRastellivanRees12}.}
\begin{IEEEeqnarray}{ll}
f_{12}\left(\xi\right) = &\left(4\xi\right)^{-\frac{\Delta_1 + \Delta_2}{2}} \Bigg[\delta_{12} + \sum_{I} 2^{\Delta_I} \textrm{C}_I \, \textrm{C}_{12}^{I} \nonumber \\
& F_{\text{bulk}}\left(\Delta_{I},\Delta_1-\Delta_2,\xi\right)\Bigg], \quad \xi \equiv \frac{\x_{12}^2}{4\z_1\z_2}. \qquad \label{StructureConstantDefect}
\end{IEEEeqnarray}
The bulk conformal blocks $F_{\text{bulk}}$ have been determined in \cite{McAvityOsborn95, LiendoRastellivanRees12} from the expression $C\left[\x_1 - \x_2,\partial_{\x_2}\right] \x_2^{-\Delta_I}$:
\begin{IEEEeqnarray}{l}
F_{\text{bulk}}\left(\Delta_{I},\delta\Delta,\xi\right) = \nonumber \\
= \xi^{\frac{\Delta_I}{2}} {_2}F_1\Big(\frac{\Delta_I + \delta\Delta}{2}, \frac{\Delta_I - \delta\Delta}{2},\Delta_I - 1;-\xi\Big), \qquad \label{DefectConformalBlocks}
\end{IEEEeqnarray}
for $\delta\Delta \equiv \Delta_1 - \Delta_2$. Dividing the (generic) dCFT two-point function \eqref{TwoPointFunctionsDefect} by the (generic) CFT two-point function \eqref{TwoPointFunctionCFT} (for $\Delta_{1,2} = L_{1,2}$) we are led to
\begin{IEEEeqnarray}{ll}
\frac{\langle\OO_{1}\left(\x_1\right)\OO_{2}\left(\x_2\right)\rangle_{\text{D5}}}{\langle\OO_{1}\left(\x_1\right)\OO_{2} \left(\x_2\right)\rangle_{\N = 4}} = \xi^{\frac{L_1 + L_2}{2}} \cdot \frac{f_{12}\left(\xi\right)}{\delta_{12}}. \qquad \label{TwoPointFunctionArbitrary2}
\end{IEEEeqnarray}
Plugging \eqref{StructureConstantDefect}--\eqref{DefectConformalBlocks} into \eqref{TwoPointFunctionArbitrary2} and concentrating on the contribution of a single protected primary operator of dimension $\Delta_I = L = 2j$, we get
\begin{IEEEeqnarray}{ll}
\frac{\langle\OO_{1}\left(\x_1\right)\OO_{2}\left(\x_2\right)\rangle_{\text{D5}}}{\langle\OO_{1}\left(\x_1\right)\OO_{2} \left(\x_2\right)\rangle_{\N = 4}} = 1 + &2^{L} \C_I \, \textrm{C}_{12}^{I} \, \xi^{j} \nonumber \\
& {_2}F_1\Big(j, j, 2j - 1; -\xi\Big), \qquad \label{OperatorProductExpansion2}
\end{IEEEeqnarray}
so that by expanding the hypergeometric around $\xi = 0$,
\begin{IEEEeqnarray}{l}
\frac{\langle\OO_{1}\left(\x_1\right)\OO_{2}\left(\x_2\right)\rangle_{\text{D5}}}{\langle\OO_{1}\left(\x_1\right)\OO_{2} \left(\x_2\right)\rangle_{\N = 4}} = 1 + 2^{L} \C_I \, \textrm{C}_{12}^{I} \, \xi^j \nonumber \\
\Bigg\{1 - \frac{j^2}{2j-1} \cdot\xi + \frac{j(j+1)^2}{4(2j-1)} \cdot\xi^2 + \ldots\Bigg\}, \qquad \label{OperatorProductExpansion3}
\end{IEEEeqnarray}
where $\C_I$ is the one-point function structure constant \eqref{OnePointFunctionStructureConstantD5} and $\textrm{C}_{12}^{I}$ is the generic three-point function structure constant. Comparing \eqref{OperatorProductExpansion3} with the strong coupling expansion \eqref{TwoPointFunctionArbitrary1} for the leading-order defect correlator of two arbitrary heavy operators and \eqref{TwoPointFunctionBMN2} for the NNLO defect correlator of two BMN chiral primaries (so that $\textrm{C}_{12}^{I} = \C_{12}^I$ is the structure constant \eqref{ThreePointFunctionStructureConstant}), we find complete agreement. In the case of two arbitrary heavy states, it would be interesting to verify the agreement of the subleading terms in \eqref{TwoPointFunctionArbitrary1}. To this end, an integral representation of the bulk-to-bulk propagator or even the Mellin transform of the amplitude could be useful \cite{RastelliZhou17a, GoncalvesItsios18}. \\
\indent The agreement of the leading-order correlator \eqref{TwoPointFunctionArbitrary1} of two arbitrary heavy operators and the OPE \eqref{OperatorProductExpansion3} implies that the value of the defect two-point function at strong coupling \eqref{TwoPointFunctionArbitrary1} will agree with its value at weak coupling whenever the heavy state is dual to a protected operator (e.g.\ for the correlation function \eqref{TwoPointFunctionBMN2}). This is guaranteed by the fact that the three-point function structure constant $\textrm{C}_{12}^{I}$ is protected and, in the large $\kappa$ limit (see \eqref{D5braneEmbedding}), the one-point function structure constant $\C_I $ agrees between weak and strong coupling. Obviously, agreement is no longer expected to hold for non-protected operators.
\section*{Acknowledgments}
\noindent We are thankful to M.\ Axenides, Z.\ Bajnok, C.\ Kristjansen and K.\ Zarembo for discussions. The research work of G.G.\ and D.Z.\ was supported by the Hellenic Foundation for Research and Innovation (HFRI) under the "First call for HFRI research projects to support faculty members and researchers and the procurement of high-cost research equipment grant" (MIS 1857, Project Number: 16519). The work of G.L.\ was supported by the National Development Research and Innovation Office (NKFIH) research grant K134946.
\appendix\section[Conventions]{Conventions \label{Appendix:Conventions}}
\noindent The equations of motion of type IIB supergravity afford a solution \cite{KimRomansvanNieuwenhuizen85} which consists of the AdS$_5\times\text{S}^5$ metric,
\begin{IEEEeqnarray}{c}
ds^2 = \frac{\ell^2}{z^2} \left(dx_0^2 + dx_1^2 + dx_2^2 + dx_3^2 + dz^2\right) + \ell^2 d\Omega_5^2, \qquad \label{MetricAdS5xS5}
\end{IEEEeqnarray}
written out here in the Poincar\'{e} coordinate system and Euclidean time. The line element of the unit 5-sphere $d\Omega_5$ takes the following $SO(3)\times SO(3)$ symmetric form:
\begin{IEEEeqnarray}{ll}
d\Omega_5^2 = d\psi^2 &+ \cos^2\psi\left(d\theta^2 + \sin^2\theta d\varphi^2\right) + \nonumber \\
& + \sin^2\psi\left(d\vartheta^2 + \sin^2\vartheta d\chi^2\right), \qquad \label{MetricS5so3so3}
\end{IEEEeqnarray}
where $\psi \in \left[0, \pi/2\right]$, $\theta,\vartheta \in \left[0, \pi\right]$, $\varphi,\chi \in \left[0, 2\pi\right)$. The solution \eqref{MetricAdS5xS5} is supported by a 4-form RR potential $\hat{C}$. The corresponding field strength $\hat{F} = d\hat{C}$ reads:
\begin{IEEEeqnarray}{c}
\hat{F}_{mnpqr} = \varepsilon_{mnpqr}, \qquad \hat{F}_{\mu\nu\rho\sigma\tau} = \varepsilon_{\mu\nu\rho\sigma\tau}. \qquad \label{RamondRamondPotentialAdS5xS5}
\end{IEEEeqnarray}
\indent The bulk-to-bulk propagator of a massive scalar field (mass $m$, scaling dimension $\Delta$) in AdS$_5$ is given by
\begin{IEEEeqnarray}{ll}
G_{\Delta}\left(x,z;y,w\right) = &\frac{\Gamma\left(\Delta\right)\eta^{\Delta}}{2^{\Delta + 1}\pi^2\Gamma\left(\Delta - 1\right)}\nonumber \\
& {_2}F_1\left(\frac{\Delta}{2},\frac{\Delta+1}{2},\nu + 1,\eta^2\right), \qquad \label{PropagatorBulkToBulk1}
\end{IEEEeqnarray}
where we have defined,
\begin{IEEEeqnarray}{c}
\eta \equiv \frac{2 z w}{z^2 + w^{2} + \left(x - y\right)^2}, \qquad \nu \equiv \sqrt{4 + m^2\ell^2}. \qquad
\end{IEEEeqnarray}
The asymptotic value of the propagator \eqref{PropagatorBulkToBulk1} near the AdS boundary ($w = 0$) becomes, for $K_z \equiv z^2 + \left(x - y\right)^2$:
\begin{IEEEeqnarray}{c}
\G_{\Delta}\left(x,z;y\right) \equiv \lim_{w\rightarrow 0} \frac{G_{\Delta}\left(x,z;y,w\right)}{w^{\Delta}} = \frac{\Delta - 1}{2\pi^2} \cdot \frac{z^{\Delta}}{K_z^{\Delta}}. \qquad \label{PropagatorBulkToBoundary}
\end{IEEEeqnarray}
\section[Chiral primary operators]{Chiral primary operators \label{Appendix:ChiralPrimaryOperators}}
\noindent The CPOs of $\N = 4$ SYM are given by symmetrized single-trace products of the theory's six scalar fields:
\begin{IEEEeqnarray}{c}
\OO^{\text{CPO}}_I\left(x\right) = \frac{1}{\sqrt{L}} \left(\frac{8\pi^2}{\lambda}\right)^{\frac{L}{2}} \Psi^{\mu_1\ldots \mu_L}_I \text{tr}\left[\Phi_{\mu_1}\ldots \Phi _{\mu_L}\right], \qquad \label{ChiralPrimaryOperators}
\end{IEEEeqnarray}
where $\Psi^{\mu_1\ldots \mu_L}_I$ are traceless symmetric tensors of $SO(6)$. The tensors $\Psi^{\mu_1\ldots \mu_L}_I$ define the S$^5$ spherical harmonics:
\begin{IEEEeqnarray}{ll}
Y_I\left(x_{\mu}\right) \equiv \Psi^{\mu_1\ldots \mu_L}_I x_{\mu_1}\ldots x_{\mu_L}, \nonumber \\
\Psi^{\mu_1\ldots \mu_L}_I\Psi^{\mu_1\ldots \mu_L}_J = \delta_{IJ}, \qquad \sum_{\mu=4}^{9} x_{\mu}^2 = 1, \qquad \label{SphericalHarmonicsSO6}
\end{IEEEeqnarray}
where $I,J$ are the corresponding quantum numbers. The overall factor in front of the CPOs \eqref{ChiralPrimaryOperators} ensures that their 2-point functions are normalized to unity \cite{LeeMinwallaRangamaniSeiberg98}. \\
\indent The scalar supergravity modes $s_I(x_m)$ that are dual to the CPOs \eqref{ChiralPrimaryOperators} have been identified \cite{KimRomansvanNieuwenhuizen85, LeeMinwallaRangamaniSeiberg98}. They are linear combinations of the scalar modes of the metric and the RR potential with $m^2\ell^2 = L(L-4)$ and $\nu = L-2$. The perturbation \eqref{FieldPerturbationIIBa}--\eqref{FieldPerturbationIIBb} can be expressed in terms of the modes $s(x_m, x_{\mu}) \equiv s_I(x_m) Y_I(x_{\mu})$ so that the vertex operators that show up in \eqref{VertexOperatorsIIB} are given by:
\begin{IEEEeqnarray}{l}
V^I_{mn} = \frac{2}{\N_L}\,\frac{1}{L+1} Y_I\left[2\ell^2\nabla_m \nabla_n - L(L - 1)\hat{g}_{mn}\right] \qquad \label{VertexOperators1} \\
V^I_{\mu\nu} = \frac{2L}{\N_L} Y_I \hat{g}_{\mu\nu}, \quad v^I_{mnpq} = \frac{\ell}{\N_L} \sqrt{\hat{g}_{\text{AdS}}} \, \varepsilon _{mnpqr} \nabla^r Y_I \qquad \label{VertexOperators2} \\
v^I_{\mu\nu\rho\sigma} = -\frac{\ell}{\N_L} \sqrt{\hat{g}_{\text{s}}} \, \varepsilon_{\mu\nu\rho\sigma\tau} Y_I \nabla^\tau, \qquad \label{VertexOperators3}
\end{IEEEeqnarray}
where the Latin indices ($m,n,p,q,r$) refer to the AdS$_5$ and the Greek indices ($\mu, \nu, \rho, \sigma, \tau$) to the S$^5$ coordinates. The normalization factor $\N_L$ is defined as
\begin{IEEEeqnarray}{l}
\N^2_{L} = \frac{N^2 L\left(L - 1\right)}{2^{L - 3} \pi^2(L + 1)^2}. \label{VertexOperators4}
\end{IEEEeqnarray}
\section[Integrals]{Integrals \label{Appendix:Integrals}}
\noindent The integrals $\I_{a,b}$ are defined in \eqref{IntegralsI} as follows:
\begin{IEEEeqnarray}{l}
\I_{a,b}\left(\kappa\right) \equiv \int\displaylimits_{0}^{\infty} du \, \frac{u^a}{\left[u^2 + (1 - \kappa u)^2\right]^b}, \qquad b > \frac{1}{2}. \qquad
\end{IEEEeqnarray}
For $j>1/2$ and $\kappa \rightarrow \infty$, we find:
\begin{IEEEeqnarray}{ll}
\I_{2j-2,2j+\frac{1}{2}} &= \kappa^{2j+1}B\big(2j,\frac{1}{2}\big) \nonumber \\ &\left\{1 + \left[\frac{3}{2} + \frac{\left(2j-3\right)\left(j-1\right)}{2\left(2j-1\right)}\right] \frac{1}{\kappa^2} + \ldots\right\} \qquad \label{IntegralI1} \\
\I_{2j-1,2j+\frac{3}{2}} &= \kappa^{2j+2}B\big(2j+1,\frac{1}{2}\big) \nonumber \\ &\left\{1 + \left[\frac{3}{2} + \frac{\left(2j-1\right)\left(j-1\right)}{4j}\right] \frac{1}{\kappa^2} + \ldots\right\} \qquad \label{IntegralI2} \\
\I_{2j-2,2j+\frac{3}{2}} &= \kappa^{2j+3}B\big(2j+1,\frac{1}{2}\big) \nonumber \\ &\left\{1 + \left[\frac{5}{2} + \frac{\left(2j-3\right)\left(j-1\right)}{4j}\right] \frac{1}{\kappa^2} + \ldots\right\} \qquad \label{IntegralI3} \\
\I_{2j,2j+\frac{5}{2}} &= \kappa^{2j+3}B\big(2j+2,\frac{1}{2}\big) \nonumber \\ &\left\{1 + \left[\frac{3}{2} + \frac{j\left(2j-1\right)}{2\left(2j+1\right)}\right] \frac{1}{\kappa^2} + \ldots\right\}. \qquad \label{IntegralI4}
\end{IEEEeqnarray}
\noindent The integrals $\J_{a,b}$ are defined as:
\begin{IEEEeqnarray}{l}
\J_{a,b} \equiv \int\displaylimits_{-\infty}^{+\infty} \frac{v^b\text{sech}^{2a+2}s\cdot ds}{\left(1 + v\tanh s\right)^b}, \qquad \label{IntegralsJ}
\end{IEEEeqnarray}
so that for $j,n = 0,1,2,\ldots$ and $v \rightarrow 0$ they are given by:
\begin{IEEEeqnarray}{l}
\J_{j+n,2j+2n+2} = \frac{\Gamma\left(\frac{1}{2}\right)}{\Gamma\left(j+n+\frac{3}{2}\right)} \nonumber \\ \sum_{m=j+n+1}^{\infty} \frac{\Gamma(m)}{\Gamma(m-j-n)} \cdot v^{2m}, \quad j+n = 0,1,\ldots \qquad \ \label{IntegralJ1} \\
\J_{j+n-1,2j+2n+1} = \frac{\Gamma\left(\frac{1}{2}\right)}{\left(j+n\right)\Gamma\left(j+n+\frac{1}{2}\right)} \nonumber \\ \sum_{m=j+n+1}^{\infty} \frac{\Gamma(m)}{\Gamma(m-j-n)} \cdot v^{2m-1}, \quad j+n = 1,2,\ldots \qquad \label{IntegralJ2} \\
\J_{j+n-1,2j+2n+2} = \frac{\Gamma\left(\frac{3}{2}\right)}{\left(j+n\right)\Gamma\left(j+n+\frac{3}{2}\right)} \sum_{m=j+n+1}^{\infty} \nonumber \\ \left(2m-1\right)\frac{\Gamma(m)}{\Gamma(m-j-n)} \cdot v^{2m}, \quad j+n = 1,2,\ldots \qquad \label{IntegralJ3}
\end{IEEEeqnarray}\vfill\null

\noindent
\bibliography{Bibliography}

\begin{thebibliography}{65}%
\makeatletter
\providecommand \@ifxundefined [1]{%
 \@ifx{#1\undefined}
}%
\providecommand \@ifnum [1]{%
 \ifnum #1\expandafter \@firstoftwo
 \else \expandafter \@secondoftwo
 \fi
}%
\providecommand \@ifx [1]{%
 \ifx #1\expandafter \@firstoftwo
 \else \expandafter \@secondoftwo
 \fi
}%
\providecommand \natexlab [1]{#1}%
\providecommand \enquote  [1]{``#1''}%
\providecommand \bibnamefont  [1]{#1}%
\providecommand \bibfnamefont [1]{#1}%
\providecommand \citenamefont [1]{#1}%
\providecommand \href@noop [0]{\@secondoftwo}%
\providecommand \href [0]{\begingroup \@sanitize@url \@href}%
\providecommand \@href[1]{\@@startlink{#1}\@@href}%
\providecommand \@@href[1]{\endgroup#1\@@endlink}%
\providecommand \@sanitize@url [0]{\catcode `\\12\catcode `\$12\catcode
  `\&12\catcode `\#12\catcode `\^12\catcode `\_12\catcode `\%12\relax}%
\providecommand \@@startlink[1]{}%
\providecommand \@@endlink[0]{}%
\providecommand \url  [0]{\begingroup\@sanitize@url \@url }%
\providecommand \@url [1]{\endgroup\@href {#1}{\urlprefix }}%
\providecommand \urlprefix  [0]{URL }%
\providecommand \Eprint [0]{\href }%
\providecommand \doibase [0]{https://doi.org/}%
\providecommand \selectlanguage [0]{\@gobble}%
\providecommand \bibinfo  [0]{\@secondoftwo}%
\providecommand \bibfield  [0]{\@secondoftwo}%
\providecommand \translation [1]{[#1]}%
\providecommand \BibitemOpen [0]{}%
\providecommand \bibitemStop [0]{}%
\providecommand \bibitemNoStop [0]{.\EOS\space}%
\providecommand \EOS [0]{\spacefactor3000\relax}%
\providecommand \BibitemShut  [1]{\csname bibitem#1\endcsname}%
\let\auto@bib@innerbib\@empty
\bibitem [{\citenamefont {Maldacena}(1998)}]{Maldacena97}%
  \BibitemOpen
  \bibfield  {author} {\bibinfo {author} {\bibfnamefont {J.~M.}\ \bibnamefont
  {Maldacena}},\ }\bibfield  {title} {\bibinfo {title} {{The large-$N$ limit of
  superconformal field theories and supergravity}},\ }\href@noop {} {\bibfield
  {journal} {\bibinfo  {journal} {Adv. Theor. Math. Phys.}\ }\textbf {\bibinfo
  {volume} {\textbf{2}}},\ \bibinfo {pages} {231} (\bibinfo {year} {1998})},\
  \Eprint {https://arxiv.org/abs/hep-th/9711200} {arXiv:hep-th/9711200
  [hep-th]} \BibitemShut {NoStop}%
\bibitem [{\citenamefont {Karch}\ and\ \citenamefont
  {Randall}(2001{\natexlab{a}})}]{KarchRandall01a}%
  \BibitemOpen
  \bibfield  {author} {\bibinfo {author} {\bibfnamefont {A.}~\bibnamefont
  {Karch}}\ and\ \bibinfo {author} {\bibfnamefont {L.}~\bibnamefont
  {Randall}},\ }\bibfield  {title} {\bibinfo {title} {{Localized gravity in
  string theory}},\ }\href {https://doi.org/10.1103/PhysRevLett.87.061601}
  {\bibfield  {journal} {\bibinfo  {journal} {Phys. Rev. Lett.}\ }\textbf
  {\bibinfo {volume} {\textbf{87}}},\ \bibinfo {pages} {061601} (\bibinfo
  {year} {2001}{\natexlab{a}})},\ \Eprint
  {https://arxiv.org/abs/hep-th/0105108} {arXiv:hep-th/0105108 [hep-th]}
  \BibitemShut {NoStop}%
\bibitem [{\citenamefont {Karch}\ and\ \citenamefont
  {Randall}(2001{\natexlab{b}})}]{KarchRandall01b}%
  \BibitemOpen
  \bibfield  {author} {\bibinfo {author} {\bibfnamefont {A.}~\bibnamefont
  {Karch}}\ and\ \bibinfo {author} {\bibfnamefont {L.}~\bibnamefont
  {Randall}},\ }\bibfield  {title} {\bibinfo {title} {{Open and closed string
  interpretation of susy CFT's on branes with boundaries}},\ }\href
  {https://doi.org/10.1088/1126-6708/2001/06/063} {\bibfield  {journal}
  {\bibinfo  {journal} {JHEP}\ }\textbf {\bibinfo {volume} {\textbf{06}}},\
  \bibinfo {pages} {063}},\ \Eprint {https://arxiv.org/abs/hep-th/0105132}
  {arXiv:hep-th/0105132 [hep-th]} \BibitemShut {NoStop}%
\bibitem [{\citenamefont {Minahan}\ and\ \citenamefont
  {Zarembo}(2003)}]{MinahanZarembo03}%
  \BibitemOpen
  \bibfield  {author} {\bibinfo {author} {\bibfnamefont {J.~A.}\ \bibnamefont
  {Minahan}}\ and\ \bibinfo {author} {\bibfnamefont {K.}~\bibnamefont
  {Zarembo}},\ }\bibfield  {title} {\bibinfo {title} {{The Bethe ansatz for
  $\mathcal{N} = 4$ super Yang-Mills}},\ }\href@noop {} {\bibfield  {journal}
  {\bibinfo  {journal} {JHEP}\ }\textbf {\bibinfo {volume} {\textbf{03}}},\
  \bibinfo {pages} {013}},\ \Eprint {https://arxiv.org/abs/hep-th/0212208}
  {arXiv:hep-th/0212208 [hep-th]} \BibitemShut {NoStop}%
\bibitem [{\citenamefont {Bena}\ \emph {et~al.}(2004)\citenamefont {Bena},
  \citenamefont {Polchinski},\ and\ \citenamefont
  {Roiban}}]{BenaPolchinskiRoiban03}%
  \BibitemOpen
  \bibfield  {author} {\bibinfo {author} {\bibfnamefont {I.}~\bibnamefont
  {Bena}}, \bibinfo {author} {\bibfnamefont {J.}~\bibnamefont {Polchinski}},\
  and\ \bibinfo {author} {\bibfnamefont {R.}~\bibnamefont {Roiban}},\
  }\bibfield  {title} {\bibinfo {title} {{Hidden symmetries of the
  AdS$_5\times\text{S}^5$ superstring}},\ }\href
  {https://doi.org/10.1103/PhysRevD.69.046002} {\bibfield  {journal} {\bibinfo
  {journal} {Phys. Rev.}\ }\textbf {\bibinfo {volume} {\textbf{D69}}},\
  \bibinfo {pages} {046002} (\bibinfo {year} {2004})},\ \Eprint
  {https://arxiv.org/abs/hep-th/0305116} {arXiv:hep-th/0305116 [hep-th]}
  \BibitemShut {NoStop}%
\bibitem [{\citenamefont {Beisert}\ \emph {et~al.}(2012)\citenamefont {Beisert}
  \emph {et~al.}}]{Beisertetal12}%
  \BibitemOpen
  \bibfield  {author} {\bibinfo {author} {\bibfnamefont {N.}~\bibnamefont
  {Beisert}} \emph {et~al.},\ }\bibfield  {title} {\bibinfo {title} {{Review of
  AdS/CFT integrability: An overview}},\ }\href
  {https://doi.org/10.1007/s11005-011-0529-2} {\bibfield  {journal} {\bibinfo
  {journal} {Lett. Math. Phys.}\ }\textbf {\bibinfo {volume} {\textbf{99}}},\
  \bibinfo {pages} {3} (\bibinfo {year} {2012})},\ \Eprint
  {https://arxiv.org/abs/1012.3982} {arXiv:1012.3982 [hep-th]} \BibitemShut
  {NoStop}%
\bibitem [{\citenamefont {de~Leeuw}\ \emph {et~al.}(2015)\citenamefont
  {de~Leeuw}, \citenamefont {Kristjansen},\ and\ \citenamefont
  {Zarembo}}]{deLeeuwKristjansenZarembo15}%
  \BibitemOpen
  \bibfield  {author} {\bibinfo {author} {\bibfnamefont {M.}~\bibnamefont
  {de~Leeuw}}, \bibinfo {author} {\bibfnamefont {C.}~\bibnamefont
  {Kristjansen}},\ and\ \bibinfo {author} {\bibfnamefont {K.}~\bibnamefont
  {Zarembo}},\ }\bibfield  {title} {\bibinfo {title} {{One-point functions in
  defect CFT and integrability}},\ }\href
  {https://doi.org/10.1007/JHEP08(2015)098} {\bibfield  {journal} {\bibinfo
  {journal} {JHEP}\ }\textbf {\bibinfo {volume} {\textbf{08}}},\ \bibinfo
  {pages} {098}},\ \Eprint {https://arxiv.org/abs/1506.06958} {arXiv:1506.06958
  [hep-th]} \BibitemShut {NoStop}%
\bibitem [{\citenamefont {Buhl-Mortensen}\ \emph
  {et~al.}(2016{\natexlab{a}})\citenamefont {Buhl-Mortensen}, \citenamefont
  {de~Leeuw}, \citenamefont {Kristjansen},\ and\ \citenamefont
  {Zarembo}}]{Buhl-MortensenLeeuwKristjansenZarembo15}%
  \BibitemOpen
  \bibfield  {author} {\bibinfo {author} {\bibfnamefont {I.}~\bibnamefont
  {Buhl-Mortensen}}, \bibinfo {author} {\bibfnamefont {M.}~\bibnamefont
  {de~Leeuw}}, \bibinfo {author} {\bibfnamefont {C.}~\bibnamefont
  {Kristjansen}},\ and\ \bibinfo {author} {\bibfnamefont {K.}~\bibnamefont
  {Zarembo}},\ }\bibfield  {title} {\bibinfo {title} {{One-point functions in
  AdS/dCFT from matrix product states}},\ }\href
  {https://doi.org/10.1007/JHEP02(2016)052} {\bibfield  {journal} {\bibinfo
  {journal} {JHEP}\ }\textbf {\bibinfo {volume} {\textbf{02}}},\ \bibinfo
  {pages} {052}},\ \Eprint {https://arxiv.org/abs/1512.02532} {arXiv:1512.02532
  [hep-th]} \BibitemShut {NoStop}%
\bibitem [{\citenamefont {de~Leeuw}\ \emph {et~al.}(2016)\citenamefont
  {de~Leeuw}, \citenamefont {Kristjansen},\ and\ \citenamefont
  {Mori}}]{deLeeuwKristjansenMori16}%
  \BibitemOpen
  \bibfield  {author} {\bibinfo {author} {\bibfnamefont {M.}~\bibnamefont
  {de~Leeuw}}, \bibinfo {author} {\bibfnamefont {C.}~\bibnamefont
  {Kristjansen}},\ and\ \bibinfo {author} {\bibfnamefont {S.}~\bibnamefont
  {Mori}},\ }\bibfield  {title} {\bibinfo {title} {{AdS/dCFT one-point
  functions of the $SU(3)$ sector}},\ }\href
  {https://doi.org/10.1016/j.physletb.2016.10.044} {\bibfield  {journal}
  {\bibinfo  {journal} {Phys. Lett.}\ }\textbf {\bibinfo {volume}
  {\textbf{B763}}},\ \bibinfo {pages} {197} (\bibinfo {year} {2016})},\ \Eprint
  {https://arxiv.org/abs/1607.03123} {arXiv:1607.03123 [hep-th]} \BibitemShut
  {NoStop}%
\bibitem [{\citenamefont {de~Leeuw}\ \emph {et~al.}(2018)\citenamefont
  {de~Leeuw}, \citenamefont {Kristjansen},\ and\ \citenamefont
  {Linardopoulos}}]{deLeeuwKristjansenLinardopoulos18a}%
  \BibitemOpen
  \bibfield  {author} {\bibinfo {author} {\bibfnamefont {M.}~\bibnamefont
  {de~Leeuw}}, \bibinfo {author} {\bibfnamefont {C.}~\bibnamefont
  {Kristjansen}},\ and\ \bibinfo {author} {\bibfnamefont {G.}~\bibnamefont
  {Linardopoulos}},\ }\bibfield  {title} {\bibinfo {title} {{Scalar one-point
  functions and matrix product states of AdS/dCFT}},\ }\href
  {https://doi.org/10.1016/j.physletb.2018.03.083} {\bibfield  {journal}
  {\bibinfo  {journal} {Phys. Lett.}\ }\textbf {\bibinfo {volume}
  {\textbf{B781}}},\ \bibinfo {pages} {238} (\bibinfo {year} {2018})},\ \Eprint
  {https://arxiv.org/abs/1802.01598} {arXiv:1802.01598 [hep-th]} \BibitemShut
  {NoStop}%
\bibitem [{\citenamefont {Kristjansen}\ \emph {et~al.}(2020)\citenamefont
  {Kristjansen}, \citenamefont {{M\"{u}ller}},\ and\ \citenamefont
  {Zarembo}}]{KristjansenMullerZarembo20a}%
  \BibitemOpen
  \bibfield  {author} {\bibinfo {author} {\bibfnamefont {C.}~\bibnamefont
  {Kristjansen}}, \bibinfo {author} {\bibfnamefont {D.}~\bibnamefont
  {{M\"{u}ller}}},\ and\ \bibinfo {author} {\bibfnamefont {K.}~\bibnamefont
  {Zarembo}},\ }\bibfield  {title} {\bibinfo {title} {{Integrable boundary
  states in D3-D5 dCFT: beyond scalars}},\ }\href
  {https://doi.org/10.1007/JHEP08(2020)103} {\bibfield  {journal} {\bibinfo
  {journal} {JHEP}\ }\textbf {\bibinfo {volume} {\textbf{08}}},\ \bibinfo
  {pages} {103}},\ \Eprint {https://arxiv.org/abs/2005.01392} {arXiv:2005.01392
  [hep-th]} \BibitemShut {NoStop}%
\bibitem [{\citenamefont {Buhl-Mortensen}\ \emph
  {et~al.}(2016{\natexlab{b}})\citenamefont {Buhl-Mortensen}, \citenamefont
  {de~Leeuw}, \citenamefont {Ipsen}, \citenamefont {Kristjansen},\ and\
  \citenamefont {Wilhelm}}]{Buhl-MortensenLeeuwIpsenKristjansenWilhelm16a}%
  \BibitemOpen
  \bibfield  {author} {\bibinfo {author} {\bibfnamefont {I.}~\bibnamefont
  {Buhl-Mortensen}}, \bibinfo {author} {\bibfnamefont {M.}~\bibnamefont
  {de~Leeuw}}, \bibinfo {author} {\bibfnamefont {A.~C.}\ \bibnamefont {Ipsen}},
  \bibinfo {author} {\bibfnamefont {C.}~\bibnamefont {Kristjansen}},\ and\
  \bibinfo {author} {\bibfnamefont {M.}~\bibnamefont {Wilhelm}},\ }\bibfield
  {title} {\bibinfo {title} {{One-loop one-point functions in gauge-gravity
  dualities with defects}},\ }\href
  {https://doi.org/10.1103/PhysRevLett.117.231603} {\bibfield  {journal}
  {\bibinfo  {journal} {Phys. Rev. Lett.}\ }\textbf {\bibinfo {volume}
  {\textbf{117}}},\ \bibinfo {pages} {231603} (\bibinfo {year}
  {2016}{\natexlab{b}})},\ \Eprint {https://arxiv.org/abs/1606.01886}
  {arXiv:1606.01886 [hep-th]} \BibitemShut {NoStop}%
\bibitem [{\citenamefont {Buhl-Mortensen}\ \emph
  {et~al.}(2017{\natexlab{a}})\citenamefont {Buhl-Mortensen}, \citenamefont
  {de~Leeuw}, \citenamefont {Ipsen}, \citenamefont {Kristjansen},\ and\
  \citenamefont {Wilhelm}}]{Buhl-MortensenLeeuwIpsenKristjansenWilhelm16c}%
  \BibitemOpen
  \bibfield  {author} {\bibinfo {author} {\bibfnamefont {I.}~\bibnamefont
  {Buhl-Mortensen}}, \bibinfo {author} {\bibfnamefont {M.}~\bibnamefont
  {de~Leeuw}}, \bibinfo {author} {\bibfnamefont {A.~C.}\ \bibnamefont {Ipsen}},
  \bibinfo {author} {\bibfnamefont {C.}~\bibnamefont {Kristjansen}},\ and\
  \bibinfo {author} {\bibfnamefont {M.}~\bibnamefont {Wilhelm}},\ }\bibfield
  {title} {\bibinfo {title} {{A quantum check of AdS/dCFT}},\ }\href
  {https://doi.org/10.1007/JHEP01(2017)098} {\bibfield  {journal} {\bibinfo
  {journal} {JHEP}\ }\textbf {\bibinfo {volume} {\textbf{01}}},\ \bibinfo
  {pages} {098}},\ \Eprint {https://arxiv.org/abs/1611.04603} {arXiv:1611.04603
  [hep-th]} \BibitemShut {NoStop}%
\bibitem [{\citenamefont {Buhl-Mortensen}\ \emph
  {et~al.}(2017{\natexlab{b}})\citenamefont {Buhl-Mortensen}, \citenamefont
  {de~Leeuw}, \citenamefont {Ipsen}, \citenamefont {Kristjansen},\ and\
  \citenamefont {Wilhelm}}]{Buhl-MortensenLeeuwIpsenKristjansenWilhelm17a}%
  \BibitemOpen
  \bibfield  {author} {\bibinfo {author} {\bibfnamefont {I.}~\bibnamefont
  {Buhl-Mortensen}}, \bibinfo {author} {\bibfnamefont {M.}~\bibnamefont
  {de~Leeuw}}, \bibinfo {author} {\bibfnamefont {A.}~\bibnamefont {Ipsen}},
  \bibinfo {author} {\bibfnamefont {C.}~\bibnamefont {Kristjansen}},\ and\
  \bibinfo {author} {\bibfnamefont {M.}~\bibnamefont {Wilhelm}},\ }\bibfield
  {title} {\bibinfo {title} {{Asymptotic one-point functions in gauge-string
  duality with defects}},\ }\href
  {https://doi.org/10.1103/PhysRevLett.119.261604} {\bibfield  {journal}
  {\bibinfo  {journal} {Phys. Rev. Lett.}\ }\textbf {\bibinfo {volume}
  {\textbf{119}}},\ \bibinfo {pages} {261604} (\bibinfo {year}
  {2017}{\natexlab{b}})},\ \Eprint {https://arxiv.org/abs/1704.07386}
  {arXiv:1704.07386 [hep-th]} \BibitemShut {NoStop}%
\bibitem [{\citenamefont {Gombor}\ and\ \citenamefont
  {Bajnok}(2020)}]{GomborBajnok20a}%
  \BibitemOpen
  \bibfield  {author} {\bibinfo {author} {\bibfnamefont {T.}~\bibnamefont
  {Gombor}}\ and\ \bibinfo {author} {\bibfnamefont {Z.}~\bibnamefont
  {Bajnok}},\ }\bibfield  {title} {\bibinfo {title} {{Boundary states,
  overlaps, nesting and bootstrapping AdS/dCFT}},\ }\href
  {https://doi.org/10.1007/JHEP10(2020)123} {\bibfield  {journal} {\bibinfo
  {journal} {JHEP}\ }\textbf {\bibinfo {volume} {\textbf{10}}},\ \bibinfo
  {pages} {123}},\ \Eprint {https://arxiv.org/abs/2004.11329} {arXiv:2004.11329
  [hep-th]} \BibitemShut {NoStop}%
\bibitem [{\citenamefont {Gombor}\ and\ \citenamefont
  {Bajnok}(2021)}]{GomborBajnok20b}%
  \BibitemOpen
  \bibfield  {author} {\bibinfo {author} {\bibfnamefont {T.}~\bibnamefont
  {Gombor}}\ and\ \bibinfo {author} {\bibfnamefont {Z.}~\bibnamefont
  {Bajnok}},\ }\bibfield  {title} {\bibinfo {title} {{Boundary state bootstrap
  and asymptotic overlaps in AdS/dCFT}},\ }\href
  {https://doi.org/10.1007/JHEP03(2021)222} {\bibfield  {journal} {\bibinfo
  {journal} {JHEP}\ }\textbf {\bibinfo {volume} {03}},\ \bibinfo {pages}
  {222}},\ \Eprint {https://arxiv.org/abs/2006.16151} {arXiv:2006.16151
  [hep-th]} \BibitemShut {NoStop}%
\bibitem [{\citenamefont {Kristjansen}\ \emph
  {et~al.}(2021{\natexlab{a}})\citenamefont {Kristjansen}, \citenamefont
  {{M\"{u}ller}},\ and\ \citenamefont {Zarembo}}]{KristjansenMullerZarembo20b}%
  \BibitemOpen
  \bibfield  {author} {\bibinfo {author} {\bibfnamefont {C.}~\bibnamefont
  {Kristjansen}}, \bibinfo {author} {\bibfnamefont {D.}~\bibnamefont
  {{M\"{u}ller}}},\ and\ \bibinfo {author} {\bibfnamefont {K.}~\bibnamefont
  {Zarembo}},\ }\bibfield  {title} {\bibinfo {title} {{Overlaps and fermionic
  dualities for integrable super spin chains}},\ }\href
  {https://doi.org/10.1007/JHEP03(2021)100} {\bibfield  {journal} {\bibinfo
  {journal} {JHEP}\ }\textbf {\bibinfo {volume} {\textbf{03}}},\ \bibinfo
  {pages} {100}},\ \Eprint {https://arxiv.org/abs/2011.12192} {arXiv:2011.12192
  [hep-th]} \BibitemShut {NoStop}%
\bibitem [{\citenamefont {Kristjansen}\ \emph
  {et~al.}(2021{\natexlab{b}})\citenamefont {Kristjansen}, \citenamefont
  {{M\"{u}ller}},\ and\ \citenamefont {Zarembo}}]{KristjansenMullerZarembo21}%
  \BibitemOpen
  \bibfield  {author} {\bibinfo {author} {\bibfnamefont {C.}~\bibnamefont
  {Kristjansen}}, \bibinfo {author} {\bibfnamefont {D.}~\bibnamefont
  {{M\"{u}ller}}},\ and\ \bibinfo {author} {\bibfnamefont {K.}~\bibnamefont
  {Zarembo}},\ }\bibfield  {title} {\bibinfo {title} {{Duality relations for
  overlaps of integrable boundary states in AdS/dCFT}},\ }\href
  {https://doi.org/10.1007/JHEP09(2021)004} {\bibfield  {journal} {\bibinfo
  {journal} {JHEP}\ }\textbf {\bibinfo {volume} {\textbf{09}}},\ \bibinfo
  {pages} {004}},\ \Eprint {https://arxiv.org/abs/2106.08116} {arXiv:2106.08116
  [hep-th]} \BibitemShut {NoStop}%
\bibitem [{\citenamefont {Dekel}\ and\ \citenamefont {Oz}(2011)}]{DekelOz11b}%
  \BibitemOpen
  \bibfield  {author} {\bibinfo {author} {\bibfnamefont {A.}~\bibnamefont
  {Dekel}}\ and\ \bibinfo {author} {\bibfnamefont {Y.}~\bibnamefont {Oz}},\
  }\bibfield  {title} {\bibinfo {title} {{Integrability of Green-Schwarz sigma
  models with boundaries}},\ }\href {https://doi.org/10.1007/JHEP08(2011)004}
  {\bibfield  {journal} {\bibinfo  {journal} {JHEP}\ }\textbf {\bibinfo
  {volume} {\textbf{08}}},\ \bibinfo {pages} {004}},\ \Eprint
  {https://arxiv.org/abs/1106.3446} {arXiv:1106.3446 [hep-th]} \BibitemShut
  {NoStop}%
\bibitem [{\citenamefont {Linardopoulos}\ and\ \citenamefont
  {Zarembo}(2021)}]{LinardopoulosZarembo21}%
  \BibitemOpen
  \bibfield  {author} {\bibinfo {author} {\bibfnamefont {G.}~\bibnamefont
  {Linardopoulos}}\ and\ \bibinfo {author} {\bibfnamefont {K.}~\bibnamefont
  {Zarembo}},\ }\bibfield  {title} {\bibinfo {title} {{String integrability of
  defect CFT and dynamical reflection matrices}},\ }\href
  {https://doi.org/10.1007/JHEP05(2021)203} {\bibfield  {journal} {\bibinfo
  {journal} {JHEP}\ }\textbf {\bibinfo {volume} {\textbf{05}}},\ \bibinfo
  {pages} {203}},\ \Eprint {https://arxiv.org/abs/2102.12381} {arXiv:2102.12381
  [hep-th]} \BibitemShut {NoStop}%
\bibitem [{\citenamefont {de~Leeuw}(2020)}]{deLeeuw19}%
  \BibitemOpen
  \bibfield  {author} {\bibinfo {author} {\bibfnamefont {M.}~\bibnamefont
  {de~Leeuw}},\ }\bibfield  {title} {\bibinfo {title} {{One-point functions in
  AdS/dCFT}},\ }\href {https://doi.org/10.1088/1751-8121/ab15fb} {\bibfield
  {journal} {\bibinfo  {journal} {J. Phys.}\ }\textbf {\bibinfo {volume}
  {\textbf{A53}}},\ \bibinfo {pages} {283001} (\bibinfo {year} {2020})},\
  \Eprint {https://arxiv.org/abs/1908.03444} {arXiv:1908.03444 [hep-th]}
  \BibitemShut {NoStop}%
\bibitem [{\citenamefont {de~Leeuw}\ \emph {et~al.}(2019)\citenamefont
  {de~Leeuw}, \citenamefont {Ipsen}, \citenamefont {Kristjansen},\ and\
  \citenamefont {Wilhelm}}]{deLeeuwIpsenKristjansenWilhelm17}%
  \BibitemOpen
  \bibfield  {author} {\bibinfo {author} {\bibfnamefont {M.}~\bibnamefont
  {de~Leeuw}}, \bibinfo {author} {\bibfnamefont {A.~C.}\ \bibnamefont {Ipsen}},
  \bibinfo {author} {\bibfnamefont {C.}~\bibnamefont {Kristjansen}},\ and\
  \bibinfo {author} {\bibfnamefont {M.}~\bibnamefont {Wilhelm}},\ }\bibfield
  {title} {\bibinfo {title} {{Introduction to integrability and one-point
  functions in $\mathcal{N} = 4$ SYM and its defect cousin}},\ }\bibfield
  {journal} {\bibinfo  {journal} {Les Houches Lect.\ Notes}\ }\textbf {\bibinfo
  {volume} {106}},\ \href {https://doi.org/10.1093/oso/9780198828150.003.0008}
  {10.1093/oso/9780198828150.003.0008} (\bibinfo {year} {2019}),\ \Eprint
  {https://arxiv.org/abs/1708.02525} {arXiv:1708.02525 [hep-th]} \BibitemShut
  {NoStop}%
\bibitem [{\citenamefont {Linardopoulos}(2020)}]{Linardopoulos20}%
  \BibitemOpen
  \bibfield  {author} {\bibinfo {author} {\bibfnamefont {G.}~\bibnamefont
  {Linardopoulos}},\ }\bibfield  {title} {\bibinfo {title} {{Solving
  holographic defects}},\ }\bibfield  {booktitle} {\emph {\bibinfo {booktitle}
  {{19th Hellenic School and Workshops on Elementary Particle Physics and
  Gravity (CORFU2019) Corfu, Greece, August 31-September 25, 2019}}},\
  }\href@noop {} {\bibfield  {journal} {\bibinfo  {journal} {PoS}\ }\textbf
  {\bibinfo {volume} {Corfu2019}},\ \bibinfo {pages} {141} (\bibinfo {year}
  {2020})},\ \Eprint {https://arxiv.org/abs/2005.02117} {arXiv:2005.02117
  [hep-th]} \BibitemShut {NoStop}%
\bibitem [{\citenamefont {de~Leeuw}\ \emph
  {et~al.}(2017{\natexlab{a}})\citenamefont {de~Leeuw}, \citenamefont
  {Kristjansen},\ and\ \citenamefont
  {Linardopoulos}}]{deLeeuwKristjansenLinardopoulos16}%
  \BibitemOpen
  \bibfield  {author} {\bibinfo {author} {\bibfnamefont {M.}~\bibnamefont
  {de~Leeuw}}, \bibinfo {author} {\bibfnamefont {C.}~\bibnamefont
  {Kristjansen}},\ and\ \bibinfo {author} {\bibfnamefont {G.}~\bibnamefont
  {Linardopoulos}},\ }\bibfield  {title} {\bibinfo {title} {{One-point
  functions of non-protected operators in the $SO(5)$ symmetric D3-D7 dCFT}},\
  }\href {https://doi.org/10.1088/1751-8121/aa714b} {\bibfield  {journal}
  {\bibinfo  {journal} {J. Phys.}\ }\textbf {\bibinfo {volume}
  {\textbf{A50}}},\ \bibinfo {pages} {254001} (\bibinfo {year}
  {2017}{\natexlab{a}})},\ \Eprint {https://arxiv.org/abs/1612.06236}
  {arXiv:1612.06236 [hep-th]} \BibitemShut {NoStop}%
\bibitem [{\citenamefont {Gimenez-Grau}\ \emph {et~al.}(2020)\citenamefont
  {Gimenez-Grau}, \citenamefont {Kristjansen}, \citenamefont {Volk},\ and\
  \citenamefont {Wilhelm}}]{GimenezGrauKristjansenVolkWilhelm19}%
  \BibitemOpen
  \bibfield  {author} {\bibinfo {author} {\bibfnamefont {A.}~\bibnamefont
  {Gimenez-Grau}}, \bibinfo {author} {\bibfnamefont {C.}~\bibnamefont
  {Kristjansen}}, \bibinfo {author} {\bibfnamefont {M.}~\bibnamefont {Volk}},\
  and\ \bibinfo {author} {\bibfnamefont {M.}~\bibnamefont {Wilhelm}},\
  }\bibfield  {title} {\bibinfo {title} {{A quantum framework for AdS/dCFT
  through fuzzy spherical harmonics on S$^4$}},\ }\href
  {https://doi.org/10.1007/JHEP04(2020)132} {\bibfield  {journal} {\bibinfo
  {journal} {JHEP}\ }\textbf {\bibinfo {volume} {\textbf{04}}},\ \bibinfo
  {pages} {132}},\ \Eprint {https://arxiv.org/abs/1912.02468} {arXiv:1912.02468
  [hep-th]} \BibitemShut {NoStop}%
\bibitem [{\citenamefont {de~Leeuw}\ \emph {et~al.}(2020)\citenamefont
  {de~Leeuw}, \citenamefont {Gombor}, \citenamefont {Kristjansen},
  \citenamefont {Linardopoulos},\ and\ \citenamefont
  {Pozsgay}}]{deLeeuwGomborKristjansenLinardopoulosPozsgay19}%
  \BibitemOpen
  \bibfield  {author} {\bibinfo {author} {\bibfnamefont {M.}~\bibnamefont
  {de~Leeuw}}, \bibinfo {author} {\bibfnamefont {T.}~\bibnamefont {Gombor}},
  \bibinfo {author} {\bibfnamefont {C.}~\bibnamefont {Kristjansen}}, \bibinfo
  {author} {\bibfnamefont {G.}~\bibnamefont {Linardopoulos}},\ and\ \bibinfo
  {author} {\bibfnamefont {B.}~\bibnamefont {Pozsgay}},\ }\bibfield  {title}
  {\bibinfo {title} {{Spin chain overlaps and the twisted Yangian}},\ }\href
  {https://doi.org/10.1007/JHEP01(2020)176} {\bibfield  {journal} {\bibinfo
  {journal} {JHEP}\ }\textbf {\bibinfo {volume} {\textbf{01}}},\ \bibinfo
  {pages} {176}},\ \Eprint {https://arxiv.org/abs/1912.09338} {arXiv:1912.09338
  [hep-th]} \BibitemShut {NoStop}%
\bibitem [{\citenamefont {Kristjansen}\ \emph {et~al.}(2022)\citenamefont
  {Kristjansen}, \citenamefont {Vu},\ and\ \citenamefont
  {Zarembo}}]{KristjansenVuZarembo21}%
  \BibitemOpen
  \bibfield  {author} {\bibinfo {author} {\bibfnamefont {C.}~\bibnamefont
  {Kristjansen}}, \bibinfo {author} {\bibfnamefont {D.-L.}\ \bibnamefont
  {Vu}},\ and\ \bibinfo {author} {\bibfnamefont {K.}~\bibnamefont {Zarembo}},\
  }\bibfield  {title} {\bibinfo {title} {{Integrable domain walls in ABJM
  theory}},\ }\href {https://doi.org/10.1007/JHEP02(2022)070} {\bibfield
  {journal} {\bibinfo  {journal} {JHEP}\ }\textbf {\bibinfo {volume}
  {\textbf{02}}},\ \bibinfo {pages} {070}},\ \Eprint
  {https://arxiv.org/abs/2112.10438} {arXiv:2112.10438 [hep-th]} \BibitemShut
  {NoStop}%
\bibitem [{\citenamefont {Gombor}\ and\ \citenamefont
  {Kristjansen}(2022)}]{GomborKristjansen22}%
  \BibitemOpen
  \bibfield  {author} {\bibinfo {author} {\bibfnamefont {T.}~\bibnamefont
  {Gombor}}\ and\ \bibinfo {author} {\bibfnamefont {C.}~\bibnamefont
  {Kristjansen}},\ }\bibfield  {title} {\bibinfo {title} {{Overlaps for matrix
  product states of arbitrary bond dimension in ABJM theory}},\ }\href
  {https://doi.org/10.1016/j.physletb.2022.137428} {\bibfield  {journal}
  {\bibinfo  {journal} {Phys. Lett.}\ }\textbf {\bibinfo {volume}
  {\textbf{B834}}},\ \bibinfo {pages} {137428} (\bibinfo {year} {2022})},\
  \Eprint {https://arxiv.org/abs/2207.06866} {arXiv:2207.06866 [hep-th]}
  \BibitemShut {NoStop}%
\bibitem [{\citenamefont {Linardopoulos}(2022)}]{Linardopoulos22}%
  \BibitemOpen
  \bibfield  {author} {\bibinfo {author} {\bibfnamefont {G.}~\bibnamefont
  {Linardopoulos}},\ }\bibfield  {title} {\bibinfo {title} {{String
  integrability of the ABJM defect}},\ }\href
  {https://doi.org/10.1007/JHEP06(2022)033} {\bibfield  {journal} {\bibinfo
  {journal} {JHEP}\ }\textbf {\bibinfo {volume} {06}},\ \bibinfo {pages}
  {033}},\ \Eprint {https://arxiv.org/abs/2202.06824} {arXiv:2202.06824
  [hep-th]} \BibitemShut {NoStop}%
\bibitem [{\citenamefont {{DeWolfe, O.}}\ \emph {et~al.}(2002)\citenamefont
  {{DeWolfe, O.}}, \citenamefont {{Freedman, D.Z.}},\ and\ \citenamefont
  {{Ooguri, H.}}}]{DeWolfeFreedmanOoguri01}%
  \BibitemOpen
  \bibfield  {author} {\bibinfo {author} {\bibnamefont {{DeWolfe, O.}}},
  \bibinfo {author} {\bibnamefont {{Freedman, D.Z.}}},\ and\ \bibinfo {author}
  {\bibnamefont {{Ooguri, H.}}},\ }\bibfield  {title} {\bibinfo {title}
  {{Holography and defect conformal field theories}},\ }\href
  {https://doi.org/10.1103/PhysRevD.66.025009} {\bibfield  {journal} {\bibinfo
  {journal} {Phys. Rev.}\ }\textbf {\bibinfo {volume} {\textbf{D66}}},\
  \bibinfo {pages} {025009} (\bibinfo {year} {2002})},\ \Eprint
  {https://arxiv.org/abs/hep-th/0111135} {arXiv:hep-th/0111135 [hep-th]}
  \BibitemShut {NoStop}%
\bibitem [{\citenamefont {de~Leeuw}\ \emph
  {et~al.}(2017{\natexlab{b}})\citenamefont {de~Leeuw}, \citenamefont {Ipsen},
  \citenamefont {Kristjansen}, \citenamefont {Vardinghus},\ and\ \citenamefont
  {Wilhelm}}]{deLeeuwIpsenKristjansenVardinghusWilhelm17}%
  \BibitemOpen
  \bibfield  {author} {\bibinfo {author} {\bibfnamefont {M.}~\bibnamefont
  {de~Leeuw}}, \bibinfo {author} {\bibfnamefont {A.~C.}\ \bibnamefont {Ipsen}},
  \bibinfo {author} {\bibfnamefont {C.}~\bibnamefont {Kristjansen}}, \bibinfo
  {author} {\bibfnamefont {K.~E.}\ \bibnamefont {Vardinghus}},\ and\ \bibinfo
  {author} {\bibfnamefont {M.}~\bibnamefont {Wilhelm}},\ }\bibfield  {title}
  {\bibinfo {title} {{Two-point functions in AdS/dCFT and the boundary
  conformal bootstrap equations}},\ }\href
  {https://doi.org/10.1007/JHEP08(2017)020} {\bibfield  {journal} {\bibinfo
  {journal} {JHEP}\ }\textbf {\bibinfo {volume} {\textbf{08}}},\ \bibinfo
  {pages} {020}},\ \Eprint {https://arxiv.org/abs/1705.03898} {arXiv:1705.03898
  [hep-th]} \BibitemShut {NoStop}%
\bibitem [{\citenamefont {Widen}(2017)}]{Widen17}%
  \BibitemOpen
  \bibfield  {author} {\bibinfo {author} {\bibfnamefont {E.}~\bibnamefont
  {Widen}},\ }\bibfield  {title} {\bibinfo {title} {{Two-point functions of
  $SU(2)$-subsector and length-two operators in dCFT}},\ }\href
  {https://doi.org/10.1016/j.physletb.2017.08.059} {\bibfield  {journal}
  {\bibinfo  {journal} {Phys. Lett.}\ }\textbf {\bibinfo {volume}
  {\textbf{B773}}},\ \bibinfo {pages} {435} (\bibinfo {year} {2017})},\ \Eprint
  {https://arxiv.org/abs/1705.08679} {arXiv:1705.08679 [hep-th]} \BibitemShut
  {NoStop}%
\bibitem [{\citenamefont {de~Leeuw}\ \emph {et~al.}(2023)\citenamefont
  {de~Leeuw}, \citenamefont {Kristjansen}, \citenamefont {Linardopoulos},\ and\
  \citenamefont {Volk}}]{deLeeuwKristjansenLinardopoulosVolk23}%
  \BibitemOpen
  \bibfield  {author} {\bibinfo {author} {\bibfnamefont {M.}~\bibnamefont
  {de~Leeuw}}, \bibinfo {author} {\bibfnamefont {C.}~\bibnamefont
  {Kristjansen}}, \bibinfo {author} {\bibfnamefont {G.}~\bibnamefont
  {Linardopoulos}},\ and\ \bibinfo {author} {\bibfnamefont {M.}~\bibnamefont
  {Volk}},\ }\bibfield  {title} {\bibinfo {title} {{B-type anomaly coefficients
  for the D3-D5 domain wall}},\ }\href@noop {} {\  (\bibinfo {year} {2023})},\
  \Eprint {https://arxiv.org/abs/2307.10946} {arXiv:2307.10946 [hep-th]}
  \BibitemShut {NoStop}%
\bibitem [{\citenamefont {Nagasaki}\ and\ \citenamefont
  {Yamaguchi}(2012)}]{NagasakiYamaguchi12}%
  \BibitemOpen
  \bibfield  {author} {\bibinfo {author} {\bibfnamefont {K.}~\bibnamefont
  {Nagasaki}}\ and\ \bibinfo {author} {\bibfnamefont {S.}~\bibnamefont
  {Yamaguchi}},\ }\bibfield  {title} {\bibinfo {title} {{Expectation values of
  chiral primary operators in holographic interface CFT}},\ }\href
  {https://doi.org/10.1103/PhysRevD.86.086004} {\bibfield  {journal} {\bibinfo
  {journal} {Phys. Rev.}\ }\textbf {\bibinfo {volume} {\textbf{D86}}},\
  \bibinfo {pages} {086004} (\bibinfo {year} {2012})},\ \Eprint
  {https://arxiv.org/abs/1205.1674} {arXiv:1205.1674 [hep-th]} \BibitemShut
  {NoStop}%
\bibitem [{\citenamefont {Kristjansen}\ \emph {et~al.}(2013)\citenamefont
  {Kristjansen}, \citenamefont {Semenoff},\ and\ \citenamefont
  {Young}}]{KristjansenSemenoffYoung12b}%
  \BibitemOpen
  \bibfield  {author} {\bibinfo {author} {\bibfnamefont {C.}~\bibnamefont
  {Kristjansen}}, \bibinfo {author} {\bibfnamefont {G.~W.}\ \bibnamefont
  {Semenoff}},\ and\ \bibinfo {author} {\bibfnamefont {D.}~\bibnamefont
  {Young}},\ }\bibfield  {title} {\bibinfo {title} {{Chiral primary one-point
  functions in the D3-D7 defect conformal field theory}},\ }\href
  {https://doi.org/10.1007/JHEP01(2013)117} {\bibfield  {journal} {\bibinfo
  {journal} {JHEP}\ }\textbf {\bibinfo {volume} {\textbf{01}}},\ \bibinfo
  {pages} {117}},\ \Eprint {https://arxiv.org/abs/1210.7015} {arXiv:1210.7015
  [hep-th]} \BibitemShut {NoStop}%
\bibitem [{\citenamefont {Robinson}\ and\ \citenamefont
  {Uhlemann}(2017)}]{RobinsonUhlemann17}%
  \BibitemOpen
  \bibfield  {author} {\bibinfo {author} {\bibfnamefont {B.}~\bibnamefont
  {Robinson}}\ and\ \bibinfo {author} {\bibfnamefont {C.~F.}\ \bibnamefont
  {Uhlemann}},\ }\bibfield  {title} {\bibinfo {title} {{Supersymmetric D3/D5
  for massive defects on curved space}},\ }\href
  {https://doi.org/10.1007/JHEP12(2017)143} {\bibfield  {journal} {\bibinfo
  {journal} {JHEP}\ }\textbf {\bibinfo {volume} {\textbf{12}}},\ \bibinfo
  {pages} {143}},\ \Eprint {https://arxiv.org/abs/1709.08650} {arXiv:1709.08650
  [hep-th]} \BibitemShut {NoStop}%
\bibitem [{\citenamefont {Wang}(2020)}]{Wang20a}%
  \BibitemOpen
  \bibfield  {author} {\bibinfo {author} {\bibfnamefont {Y.}~\bibnamefont
  {Wang}},\ }\bibfield  {title} {\bibinfo {title} {{Taming defects in $
  \mathcal{N} $ = 4 super-Yang-Mills}},\ }\href
  {https://doi.org/10.1007/JHEP08(2020)021} {\bibfield  {journal} {\bibinfo
  {journal} {JHEP}\ }\textbf {\bibinfo {volume} {\textbf{08}}},\ \bibinfo
  {pages} {021}},\ \Eprint {https://arxiv.org/abs/2003.11016} {arXiv:2003.11016
  [hep-th]} \BibitemShut {NoStop}%
\bibitem [{\citenamefont {Komatsu}\ and\ \citenamefont
  {Wang}(2020)}]{KomatsuWang20}%
  \BibitemOpen
  \bibfield  {author} {\bibinfo {author} {\bibfnamefont {S.}~\bibnamefont
  {Komatsu}}\ and\ \bibinfo {author} {\bibfnamefont {Y.}~\bibnamefont {Wang}},\
  }\bibfield  {title} {\bibinfo {title} {{Non-perturbative defect one-point
  functions in planar $\mathcal{N}=4$ super-Yang-Mills}},\ }\href
  {https://doi.org/10.1016/j.nuclphysb.2020.115120} {\bibfield  {journal}
  {\bibinfo  {journal} {Nucl. Phys.}\ }\textbf {\bibinfo {volume}
  {\textbf{B958}}},\ \bibinfo {pages} {115120} (\bibinfo {year} {2020})},\
  \Eprint {https://arxiv.org/abs/2004.09514} {arXiv:2004.09514 [hep-th]}
  \BibitemShut {NoStop}%
\bibitem [{\citenamefont {Beccaria}\ and\ \citenamefont
  {Cabo-Bizet}(2023)}]{BeccariaCaboBizet23}%
  \BibitemOpen
  \bibfield  {author} {\bibinfo {author} {\bibfnamefont {M.}~\bibnamefont
  {Beccaria}}\ and\ \bibinfo {author} {\bibfnamefont {A.}~\bibnamefont
  {Cabo-Bizet}},\ }\bibfield  {title} {\bibinfo {title} {{$1/N$ expansion of
  the D3-D5 defect CFT at strong coupling}},\ }\href
  {https://doi.org/10.1007/JHEP02(2023)208} {\bibfield  {journal} {\bibinfo
  {journal} {JHEP}\ }\textbf {\bibinfo {volume} {\textbf{02}}},\ \bibinfo
  {pages} {208}},\ \Eprint {https://arxiv.org/abs/2212.12415} {arXiv:2212.12415
  [hep-th]} \BibitemShut {NoStop}%
\bibitem [{\citenamefont {Bak}\ \emph {et~al.}(2011)\citenamefont {Bak},
  \citenamefont {Chen},\ and\ \citenamefont {Wu}}]{BakChenWu11}%
  \BibitemOpen
  \bibfield  {author} {\bibinfo {author} {\bibfnamefont {D.}~\bibnamefont
  {Bak}}, \bibinfo {author} {\bibfnamefont {B.}~\bibnamefont {Chen}},\ and\
  \bibinfo {author} {\bibfnamefont {J.-B.}\ \bibnamefont {Wu}},\ }\bibfield
  {title} {\bibinfo {title} {{Holographic correlation functions for open
  strings and branes}},\ }\href {https://doi.org/10.1007/JHEP06(2011)014}
  {\bibfield  {journal} {\bibinfo  {journal} {JHEP}\ }\textbf {\bibinfo
  {volume} {\textbf{06}}},\ \bibinfo {pages} {014}},\ \Eprint
  {https://arxiv.org/abs/1103.2024} {arXiv:1103.2024 [hep-th]} \BibitemShut
  {NoStop}%
\bibitem [{\citenamefont {Bissi}\ \emph {et~al.}(2011)\citenamefont {Bissi},
  \citenamefont {Kristjansen}, \citenamefont {Young},\ and\ \citenamefont
  {Zoubos}}]{BissiKristjansenYoungZoubos11}%
  \BibitemOpen
  \bibfield  {author} {\bibinfo {author} {\bibfnamefont {A.}~\bibnamefont
  {Bissi}}, \bibinfo {author} {\bibfnamefont {C.}~\bibnamefont {Kristjansen}},
  \bibinfo {author} {\bibfnamefont {D.}~\bibnamefont {Young}},\ and\ \bibinfo
  {author} {\bibfnamefont {K.}~\bibnamefont {Zoubos}},\ }\bibfield  {title}
  {\bibinfo {title} {{Holographic three-point functions of giant gravitons}},\
  }\href {https://doi.org/10.1007/JHEP06(2011)085} {\bibfield  {journal}
  {\bibinfo  {journal} {JHEP}\ }\textbf {\bibinfo {volume} {\textbf{06}}},\
  \bibinfo {pages} {085}},\ \Eprint {https://arxiv.org/abs/1103.4079}
  {arXiv:1103.4079 [hep-th]} \BibitemShut {NoStop}%
\bibitem [{\citenamefont {Caputa}\ \emph {et~al.}(2012)\citenamefont {Caputa},
  \citenamefont {de~Mello~Koch},\ and\ \citenamefont
  {Zoubos}}]{CaputaMelloKochZoubos12}%
  \BibitemOpen
  \bibfield  {author} {\bibinfo {author} {\bibfnamefont {P.}~\bibnamefont
  {Caputa}}, \bibinfo {author} {\bibfnamefont {R.}~\bibnamefont
  {de~Mello~Koch}},\ and\ \bibinfo {author} {\bibfnamefont {K.}~\bibnamefont
  {Zoubos}},\ }\bibfield  {title} {\bibinfo {title} {{Extremal versus
  non-extremal correlators with giant gravitons}},\ }\href
  {https://doi.org/10.1007/JHEP08(2012)143} {\bibfield  {journal} {\bibinfo
  {journal} {JHEP}\ }\textbf {\bibinfo {volume} {\textbf{08}}},\ \bibinfo
  {pages} {143}},\ \Eprint {https://arxiv.org/abs/1204.4172} {arXiv:1204.4172
  [hep-th]} \BibitemShut {NoStop}%
\bibitem [{\citenamefont {Hirano}\ \emph {et~al.}(2012)\citenamefont {Hirano},
  \citenamefont {Kristjansen},\ and\ \citenamefont
  {Young}}]{HiranoKristjansenYoung12}%
  \BibitemOpen
  \bibfield  {author} {\bibinfo {author} {\bibfnamefont {S.}~\bibnamefont
  {Hirano}}, \bibinfo {author} {\bibfnamefont {C.}~\bibnamefont
  {Kristjansen}},\ and\ \bibinfo {author} {\bibfnamefont {D.}~\bibnamefont
  {Young}},\ }\bibfield  {title} {\bibinfo {title} {{Giant gravitons on AdS$_4
  \times \mathbb{CP}^3$ and their holographic three-point functions}},\ }\href
  {https://doi.org/10.1007/JHEP07(2012)006} {\bibfield  {journal} {\bibinfo
  {journal} {JHEP}\ }\textbf {\bibinfo {volume} {\textbf{07}}},\ \bibinfo
  {pages} {006}},\ \Eprint {https://arxiv.org/abs/1205.1959} {arXiv:1205.1959
  [hep-th]} \BibitemShut {NoStop}%
\bibitem [{\citenamefont {Lin}(2012)}]{Lin12}%
  \BibitemOpen
  \bibfield  {author} {\bibinfo {author} {\bibfnamefont {H.}~\bibnamefont
  {Lin}},\ }\bibfield  {title} {\bibinfo {title} {{Giant gravitons and
  correlators}},\ }\href {https://doi.org/10.1007/JHEP12(2012)011} {\bibfield
  {journal} {\bibinfo  {journal} {JHEP}\ }\textbf {\bibinfo {volume}
  {\textbf{12}}},\ \bibinfo {pages} {011}},\ \Eprint
  {https://arxiv.org/abs/1209.6624} {arXiv:1209.6624 [hep-th]} \BibitemShut
  {NoStop}%
\bibitem [{\citenamefont {Kristjansen}\ \emph {et~al.}(2015)\citenamefont
  {Kristjansen}, \citenamefont {Mori},\ and\ \citenamefont
  {Young}}]{KristjansenMoriYoung15}%
  \BibitemOpen
  \bibfield  {author} {\bibinfo {author} {\bibfnamefont {C.}~\bibnamefont
  {Kristjansen}}, \bibinfo {author} {\bibfnamefont {S.}~\bibnamefont {Mori}},\
  and\ \bibinfo {author} {\bibfnamefont {D.}~\bibnamefont {Young}},\ }\bibfield
   {title} {\bibinfo {title} {{On the regularization of extremal three-point
  functions involving giant gravitons}},\ }\href
  {https://doi.org/10.1016/j.physletb.2015.09.056} {\bibfield  {journal}
  {\bibinfo  {journal} {Phys. Lett.}\ }\textbf {\bibinfo {volume}
  {\textbf{B750}}},\ \bibinfo {pages} {379} (\bibinfo {year} {2015})},\ \Eprint
  {https://arxiv.org/abs/1507.03965} {arXiv:1507.03965 [hep-th]} \BibitemShut
  {NoStop}%
\bibitem [{\citenamefont {Berenstein}\ \emph {et~al.}(1999)\citenamefont
  {Berenstein}, \citenamefont {Corrado}, \citenamefont {Fischler},\ and\
  \citenamefont {Maldacena}}]{BerensteinCorradoFischlerMaldacena99}%
  \BibitemOpen
  \bibfield  {author} {\bibinfo {author} {\bibfnamefont {D.~E.}\ \bibnamefont
  {Berenstein}}, \bibinfo {author} {\bibfnamefont {R.}~\bibnamefont {Corrado}},
  \bibinfo {author} {\bibfnamefont {W.}~\bibnamefont {Fischler}},\ and\
  \bibinfo {author} {\bibfnamefont {J.~M.}\ \bibnamefont {Maldacena}},\
  }\bibfield  {title} {\bibinfo {title} {{The operator product expansion for
  Wilson loops and surfaces in the large-$N$ limit}},\ }\href
  {https://doi.org/10.1103/PhysRevD.59.105023} {\bibfield  {journal} {\bibinfo
  {journal} {Phys. Rev.}\ }\textbf {\bibinfo {volume} {\textbf{D59}}},\
  \bibinfo {pages} {105023} (\bibinfo {year} {1999})},\ \Eprint
  {https://arxiv.org/abs/hep-th/9809188} {arXiv:hep-th/9809188 [hep-th]}
  \BibitemShut {NoStop}%
\bibitem [{\citenamefont {Zarembo}(2010)}]{Zarembo10c}%
  \BibitemOpen
  \bibfield  {author} {\bibinfo {author} {\bibfnamefont {K.}~\bibnamefont
  {Zarembo}},\ }\bibfield  {title} {\bibinfo {title} {{Holographic three-point
  functions of semiclassical states}},\ }\href
  {https://doi.org/10.1007/JHEP09(2010)030} {\bibfield  {journal} {\bibinfo
  {journal} {JHEP}\ }\textbf {\bibinfo {volume} {\textbf{09}}},\ \bibinfo
  {pages} {030}},\ \Eprint {https://arxiv.org/abs/1008.1059} {arXiv:1008.1059
  [hep-th]} \BibitemShut {NoStop}%
\bibitem [{\citenamefont {Costa}\ \emph {et~al.}(2010)\citenamefont {Costa},
  \citenamefont {Monteiro}, \citenamefont {Santos},\ and\ \citenamefont
  {Zoakos}}]{CostaMonteiroSantosZoakos10}%
  \BibitemOpen
  \bibfield  {author} {\bibinfo {author} {\bibfnamefont {M.~S.}\ \bibnamefont
  {Costa}}, \bibinfo {author} {\bibfnamefont {R.}~\bibnamefont {Monteiro}},
  \bibinfo {author} {\bibfnamefont {J.~E.}\ \bibnamefont {Santos}},\ and\
  \bibinfo {author} {\bibfnamefont {D.}~\bibnamefont {Zoakos}},\ }\bibfield
  {title} {\bibinfo {title} {{On three-point correlation functions in the
  gauge/gravity duality}},\ }\href {https://doi.org/10.1007/JHEP11(2010)141}
  {\bibfield  {journal} {\bibinfo  {journal} {JHEP}\ }\textbf {\bibinfo
  {volume} {11}},\ \bibinfo {pages} {141}},\ \Eprint
  {https://arxiv.org/abs/1008.1070} {arXiv:1008.1070 [hep-th]} \BibitemShut
  {NoStop}%
\bibitem [{\citenamefont {Tsuji}(2007)}]{Tsuji07}%
  \BibitemOpen
  \bibfield  {author} {\bibinfo {author} {\bibfnamefont {A.}~\bibnamefont
  {Tsuji}},\ }\bibfield  {title} {\bibinfo {title} {{Holography of Wilson loop
  correlator and spinning strings}},\ }\href
  {https://doi.org/10.1143/PTP.117.557} {\bibfield  {journal} {\bibinfo
  {journal} {Prog. Theor. Phys.}\ }\textbf {\bibinfo {volume} {117}},\ \bibinfo
  {pages} {\textbf{557}} (\bibinfo {year} {2007})},\ \Eprint
  {https://arxiv.org/abs/hep-th/0606030} {arXiv:hep-th/0606030 [hep-th]}
  \BibitemShut {NoStop}%
\bibitem [{\citenamefont {Janik}\ \emph {et~al.}(2010)\citenamefont {Janik},
  \citenamefont {Surowka},\ and\ \citenamefont
  {Wereszczynski}}]{JanikSurowkaWereszczynski10}%
  \BibitemOpen
  \bibfield  {author} {\bibinfo {author} {\bibfnamefont {R.~A.}\ \bibnamefont
  {Janik}}, \bibinfo {author} {\bibfnamefont {P.}~\bibnamefont {Surowka}},\
  and\ \bibinfo {author} {\bibfnamefont {A.}~\bibnamefont {Wereszczynski}},\
  }\bibfield  {title} {\bibinfo {title} {{On correlation functions of operators
  dual to classical spinning string states}},\ }\href
  {https://doi.org/10.1007/JHEP05(2010)030} {\bibfield  {journal} {\bibinfo
  {journal} {JHEP}\ }\textbf {\bibinfo {volume} {05}},\ \bibinfo {pages}
  {030}},\ \Eprint {https://arxiv.org/abs/1002.4613} {arXiv:1002.4613 [hep-th]}
  \BibitemShut {NoStop}%
\bibitem [{\citenamefont {Lee}\ \emph {et~al.}(1998)\citenamefont {Lee},
  \citenamefont {Minwalla}, \citenamefont {Rangamani},\ and\ \citenamefont
  {Seiberg}}]{LeeMinwallaRangamaniSeiberg98}%
  \BibitemOpen
  \bibfield  {author} {\bibinfo {author} {\bibfnamefont {S.}~\bibnamefont
  {Lee}}, \bibinfo {author} {\bibfnamefont {S.}~\bibnamefont {Minwalla}},
  \bibinfo {author} {\bibfnamefont {M.}~\bibnamefont {Rangamani}},\ and\
  \bibinfo {author} {\bibfnamefont {N.}~\bibnamefont {Seiberg}},\ }\bibfield
  {title} {\bibinfo {title} {{Three point functions of chiral operators in $D =
  4$, $\mathcal{N} = 4$ SYM at large $N$}},\ }\href
  {https://doi.org/10.4310/ATMP.1998.v2.n4.a1} {\bibfield  {journal} {\bibinfo
  {journal} {Adv. Theor. Math. Phys.}\ }\textbf {\bibinfo {volume}
  {\textbf{2}}},\ \bibinfo {pages} {697} (\bibinfo {year} {1998})},\ \Eprint
  {https://arxiv.org/abs/hep-th/9806074} {arXiv:hep-th/9806074 [hep-th]}
  \BibitemShut {NoStop}%
\bibitem [{\citenamefont {Roiban}\ and\ \citenamefont
  {Tseytlin}(2010)}]{RoibanTseytlin10}%
  \BibitemOpen
  \bibfield  {author} {\bibinfo {author} {\bibfnamefont {R.}~\bibnamefont
  {Roiban}}\ and\ \bibinfo {author} {\bibfnamefont {A.~A.}\ \bibnamefont
  {Tseytlin}},\ }\bibfield  {title} {\bibinfo {title} {{On semiclassical
  computation of three-point functions of closed string vertex operators in
  AdS$_5\times\text{S}^5$}},\ }\href
  {https://doi.org/10.1103/PhysRevD.82.106011} {\bibfield  {journal} {\bibinfo
  {journal} {Phys. Rev.}\ }\textbf {\bibinfo {volume} {\textbf{D82}}},\
  \bibinfo {pages} {106011} (\bibinfo {year} {2010})},\ \Eprint
  {https://arxiv.org/abs/1008.4921} {arXiv:1008.4921 [hep-th]} \BibitemShut
  {NoStop}%
\bibitem [{\citenamefont {Georgiou}(2011{\natexlab{a}})}]{Georgiou10}%
  \BibitemOpen
  \bibfield  {author} {\bibinfo {author} {\bibfnamefont {G.}~\bibnamefont
  {Georgiou}},\ }\bibfield  {title} {\bibinfo {title} {{Two and three-point
  correlators of operators dual to folded string solutions at strong
  coupling}},\ }\href {https://doi.org/10.1007/JHEP02(2011)046} {\bibfield
  {journal} {\bibinfo  {journal} {JHEP}\ }\textbf {\bibinfo {volume}
  {\textbf{02}}},\ \bibinfo {pages} {046}},\ \Eprint
  {https://arxiv.org/abs/1011.5181} {arXiv:1011.5181 [hep-th]} \BibitemShut
  {NoStop}%
\bibitem [{\citenamefont {Georgiou}(2011{\natexlab{b}})}]{Georgiou11}%
  \BibitemOpen
  \bibfield  {author} {\bibinfo {author} {\bibfnamefont {G.}~\bibnamefont
  {Georgiou}},\ }\bibfield  {title} {\bibinfo {title} {{$SL(2)$ sector:
  Weak/strong coupling agreement of three-point correlators}},\ }\href
  {https://doi.org/10.1007/JHEP09(2011)132} {\bibfield  {journal} {\bibinfo
  {journal} {JHEP}\ }\textbf {\bibinfo {volume} {\textbf{09}}},\ \bibinfo
  {pages} {132}},\ \Eprint {https://arxiv.org/abs/1107.1850} {arXiv:1107.1850
  [hep-th]} \BibitemShut {NoStop}%
\bibitem [{\citenamefont {Georgiou}\ \emph {et~al.}(2013)\citenamefont
  {Georgiou}, \citenamefont {Lee},\ and\ \citenamefont
  {Park}}]{GeorgiouLeePark13}%
  \BibitemOpen
  \bibfield  {author} {\bibinfo {author} {\bibfnamefont {G.}~\bibnamefont
  {Georgiou}}, \bibinfo {author} {\bibfnamefont {B.-H.}\ \bibnamefont {Lee}},\
  and\ \bibinfo {author} {\bibfnamefont {C.}~\bibnamefont {Park}},\ }\bibfield
  {title} {\bibinfo {title} {{Correlators of massive string states with
  conserved currents}},\ }\href {https://doi.org/10.1007/JHEP03(2013)167}
  {\bibfield  {journal} {\bibinfo  {journal} {JHEP}\ }\textbf {\bibinfo
  {volume} {\textbf{03}}},\ \bibinfo {pages} {167}},\ \Eprint
  {https://arxiv.org/abs/1301.5092} {arXiv:1301.5092 [hep-th]} \BibitemShut
  {NoStop}%
\bibitem [{\citenamefont {Bajnok}\ \emph {et~al.}(2014)\citenamefont {Bajnok},
  \citenamefont {Janik},\ and\ \citenamefont
  {Wereszczy\'nski}}]{BajnokJanikWereszczynski14}%
  \BibitemOpen
  \bibfield  {author} {\bibinfo {author} {\bibfnamefont {Z.}~\bibnamefont
  {Bajnok}}, \bibinfo {author} {\bibfnamefont {R.~A.}\ \bibnamefont {Janik}},\
  and\ \bibinfo {author} {\bibfnamefont {A.}~\bibnamefont {Wereszczy\'nski}},\
  }\bibfield  {title} {\bibinfo {title} {{HHL correlators, orbit averaging and
  form factors}},\ }\href {https://doi.org/10.1007/JHEP09(2014)050} {\bibfield
  {journal} {\bibinfo  {journal} {JHEP}\ }\textbf {\bibinfo {volume}
  {\textbf{09}}},\ \bibinfo {pages} {050}},\ \Eprint
  {https://arxiv.org/abs/1404.4556} {arXiv:1404.4556 [hep-th]} \BibitemShut
  {NoStop}%
\bibitem [{\citenamefont {Bajnok}\ and\ \citenamefont
  {Janik}(2017)}]{BajnokJanik17a}%
  \BibitemOpen
  \bibfield  {author} {\bibinfo {author} {\bibfnamefont {Z.}~\bibnamefont
  {Bajnok}}\ and\ \bibinfo {author} {\bibfnamefont {R.~A.}\ \bibnamefont
  {Janik}},\ }\bibfield  {title} {\bibinfo {title} {{Classical limit of
  diagonal form factors and HHL correlators}},\ }\href
  {https://doi.org/10.1007/JHEP01(2017)063} {\bibfield  {journal} {\bibinfo
  {journal} {JHEP}\ }\textbf {\bibinfo {volume} {\textbf{01}}},\ \bibinfo
  {pages} {063}},\ \Eprint {https://arxiv.org/abs/1607.02830} {arXiv:1607.02830
  [hep-th]} \BibitemShut {NoStop}%
\bibitem [{\citenamefont {Gubser}\ \emph {et~al.}(1998)\citenamefont {Gubser},
  \citenamefont {Klebanov},\ and\ \citenamefont
  {Polyakov}}]{GubserKlebanovPolyakov98}%
  \BibitemOpen
  \bibfield  {author} {\bibinfo {author} {\bibfnamefont {S.~S.}\ \bibnamefont
  {Gubser}}, \bibinfo {author} {\bibfnamefont {I.~R.}\ \bibnamefont
  {Klebanov}},\ and\ \bibinfo {author} {\bibfnamefont {A.~M.}\ \bibnamefont
  {Polyakov}},\ }\bibfield  {title} {\bibinfo {title} {{Gauge theory
  correlators from non-critical string theory}},\ }\href
  {https://doi.org/10.1016/S0370-2693(98)00377-3} {\bibfield  {journal}
  {\bibinfo  {journal} {Phys. Lett.}\ }\textbf {\bibinfo {volume}
  {\textbf{B428}}},\ \bibinfo {pages} {105} (\bibinfo {year} {1998})},\ \Eprint
  {https://arxiv.org/abs/hep-th/9802109} {arXiv:hep-th/9802109 [hep-th]}
  \BibitemShut {NoStop}%
\bibitem [{\citenamefont {Witten}(1998)}]{Witten98a}%
  \BibitemOpen
  \bibfield  {author} {\bibinfo {author} {\bibfnamefont {E.}~\bibnamefont
  {Witten}},\ }\bibfield  {title} {\bibinfo {title} {{Anti-de Sitter space and
  holography}},\ }\href@noop {} {\bibfield  {journal} {\bibinfo  {journal}
  {Adv. Theor. Math. Phys.}\ }\textbf {\bibinfo {volume} {\textbf{2}}},\
  \bibinfo {pages} {253} (\bibinfo {year} {1998})},\ \Eprint
  {https://arxiv.org/abs/hep-th/9802150} {arXiv:hep-th/9802150 [hep-th]}
  \BibitemShut {NoStop}%
\bibitem [{\citenamefont {Kazama}\ and\ \citenamefont
  {Komatsu}(2012)}]{KazamaKomatsu11}%
  \BibitemOpen
  \bibfield  {author} {\bibinfo {author} {\bibfnamefont {Y.}~\bibnamefont
  {Kazama}}\ and\ \bibinfo {author} {\bibfnamefont {S.}~\bibnamefont
  {Komatsu}},\ }\bibfield  {title} {\bibinfo {title} {{On holographic three
  point functions for GKP strings from integrability}},\ }\href
  {https://doi.org/10.1007/JHEP01(2012)110} {\bibfield  {journal} {\bibinfo
  {journal} {JHEP}\ }\textbf {\bibinfo {volume} {\textbf{01}}},\ \bibinfo
  {pages} {110}},\ \bibinfo {note} {[Erratum: JHEP \textbf{06} (2012) 150]},\
  \Eprint {https://arxiv.org/abs/1110.3949} {arXiv:1110.3949 [hep-th]}
  \BibitemShut {NoStop}%
\bibitem [{\citenamefont {McAvity}\ and\ \citenamefont
  {Osborn}(1995)}]{McAvityOsborn95}%
  \BibitemOpen
  \bibfield  {author} {\bibinfo {author} {\bibfnamefont {D.~M.}\ \bibnamefont
  {McAvity}}\ and\ \bibinfo {author} {\bibfnamefont {H.}~\bibnamefont
  {Osborn}},\ }\bibfield  {title} {\bibinfo {title} {{Conformal field theories
  near a boundary in general dimensions}},\ }\href
  {https://doi.org/10.1016/0550-3213(95)00476-9} {\bibfield  {journal}
  {\bibinfo  {journal} {Nucl. Phys.}\ }\textbf {\bibinfo {volume}
  {\textbf{B455}}},\ \bibinfo {pages} {522} (\bibinfo {year} {1995})},\ \Eprint
  {https://arxiv.org/abs/cond-mat/9505127} {arXiv:cond-mat/9505127 [cond-mat]}
  \BibitemShut {NoStop}%
\bibitem [{\citenamefont {Liendo}\ \emph {et~al.}(2013)\citenamefont {Liendo},
  \citenamefont {Rastelli},\ and\ \citenamefont {van
  Rees}}]{LiendoRastellivanRees12}%
  \BibitemOpen
  \bibfield  {author} {\bibinfo {author} {\bibfnamefont {P.}~\bibnamefont
  {Liendo}}, \bibinfo {author} {\bibfnamefont {L.}~\bibnamefont {Rastelli}},\
  and\ \bibinfo {author} {\bibfnamefont {B.~C.}\ \bibnamefont {van Rees}},\
  }\bibfield  {title} {\bibinfo {title} {{The bootstrap program for boundary
  CFT$_d$}},\ }\href {https://doi.org/10.1007/JHEP07(2013)113} {\bibfield
  {journal} {\bibinfo  {journal} {JHEP}\ }\textbf {\bibinfo {volume}
  {\textbf{07}}},\ \bibinfo {pages} {113}},\ \Eprint
  {https://arxiv.org/abs/1210.4258} {arXiv:1210.4258 [hep-th]} \BibitemShut
  {NoStop}%
\bibitem [{\citenamefont {Rastelli}\ and\ \citenamefont
  {Zhou}(2017)}]{RastelliZhou17a}%
  \BibitemOpen
  \bibfield  {author} {\bibinfo {author} {\bibfnamefont {L.}~\bibnamefont
  {Rastelli}}\ and\ \bibinfo {author} {\bibfnamefont {X.}~\bibnamefont
  {Zhou}},\ }\bibfield  {title} {\bibinfo {title} {{The Mellin formalism for
  boundary CFT$_d$}},\ }\href {https://doi.org/10.1007/JHEP10(2017)146}
  {\bibfield  {journal} {\bibinfo  {journal} {JHEP}\ }\textbf {\bibinfo
  {volume} {\textbf{10}}},\ \bibinfo {pages} {146}},\ \Eprint
  {https://arxiv.org/abs/1705.05362} {arXiv:1705.05362 [hep-th]} \BibitemShut
  {NoStop}%
\bibitem [{\citenamefont {Goncalves}\ and\ \citenamefont
  {Itsios}(2018)}]{GoncalvesItsios18}%
  \BibitemOpen
  \bibfield  {author} {\bibinfo {author} {\bibfnamefont {V.}~\bibnamefont
  {Goncalves}}\ and\ \bibinfo {author} {\bibfnamefont {G.}~\bibnamefont
  {Itsios}},\ }\bibfield  {title} {\bibinfo {title} {{A note on defect Mellin
  amplitudes}},\ }\href@noop {} {\  (\bibinfo {year} {2018})},\ \Eprint
  {https://arxiv.org/abs/1803.06721} {arXiv:1803.06721 [hep-th]} \BibitemShut
  {NoStop}%
\bibitem [{\citenamefont {Kim}\ \emph {et~al.}(1985)\citenamefont {Kim},
  \citenamefont {Romans},\ and\ \citenamefont {van
  Nieuwenhuizen}}]{KimRomansvanNieuwenhuizen85}%
  \BibitemOpen
  \bibfield  {author} {\bibinfo {author} {\bibfnamefont {H.~J.}\ \bibnamefont
  {Kim}}, \bibinfo {author} {\bibfnamefont {L.~J.}\ \bibnamefont {Romans}},\
  and\ \bibinfo {author} {\bibfnamefont {P.}~\bibnamefont {van
  Nieuwenhuizen}},\ }\bibfield  {title} {\bibinfo {title} {{The mass spectrum
  of chiral $\mathcal{N} = 2$, $D = 10$ supergravity on S$^5$}},\ }\href
  {https://doi.org/10.1103/PhysRevD.32.389} {\bibfield  {journal} {\bibinfo
  {journal} {Phys. Rev.}\ }\textbf {\bibinfo {volume} {\textbf{D32}}},\
  \bibinfo {pages} {389} (\bibinfo {year} {1985})}\BibitemShut {NoStop}%
\end{thebibliography}%

\end{document}